\documentclass[journal,hideappendix]{vgtc}        % final (journal style) without appendices

\usepackage{amsmath}
\usepackage{makecell}
\usepackage{balance}
\usepackage{bbm}

\onlineid{1291}

\vgtccategory{Research}

\vgtcpapertype{Applications}

\title{\systemname: Visual Analytics for Enhancing the Eligibility Criteria Design of Clinical Trials}

\author{%
  \authororcid{Rui Sheng}{0000-0001-9321-6756},
  \authororcid{Xingbo Wang}{0000-0001-5693-1128}, 
  \authororcid{Jiachen Wang}{0000-0001-9630-9958}, 
  \authororcid{Xiaofu Jin}{0000-0002-7239-3769}, 
  \authororcid{Zhonghua Sheng}{0009-0004-6020-0175}, \\
  \authororcid{Zhenxing Xu}{}, 
  \authororcid{Suraj Rajendran}{0000-0002-8149-0157}, 
  \authororcid{Huamin Qu}{0000-0002-3344-9694}, 
  and \authororcid{Fei Wang}{0000-0001-9459-9461}
}

\authorfooter{
  \item
  	R. Sheng, X. Jin, Z. Sheng, and H. Qu are with the Hong Kong
University of Science and Technology.
  	E-mail: \{rshengac, xjinao, szh\}@connect.ust.hk, huamin@cse.ust.hk.
  \item
  	Z. Xu, S. Rajendran, and F. Wang is with Cornell University.
  	E-mail: \{zhx2005, sur4002, few2001\}@med.cornell.edu.

  \item X. Wang is with Bosch.
  	E-mail: wangxbzb@gmail.com.

  \item X. Wang and F. Wang are the co-corresponding authors.
}

\abstract{Eligibility criteria play a critical role in clinical trials by determining the target patient population, which significantly influences the outcomes of medical interventions.
However, current approaches for designing eligibility criteria have limitations to support interactive exploration of the large space of eligibility criteria. They also ignore incorporating detailed characteristics from the original electronic health record (EHR) data for criteria refinement. 
To address these limitations, we proposed \systemname, a visual analytics system integrating a novel workflow, which can empower clinicians to iteratively explore the vast space of eligibility criteria through knowledge-driven and outcome-driven approaches.
\systemname supports history-tracking to help clinicians trace the evolution of their adjustments and decisions when exploring various forms of data (\ie, eligibility criteria, outcome metrics, and detailed characteristics of original EHR data) through these two approaches.
This feature can help clinicians comprehend the impact of eligibility criteria on outcome metrics and patient characteristics, which facilitates systematic refinement of eligibility criteria.
Using a real-world dataset, we demonstrated the effectiveness of \systemname in providing insights into designing eligibility criteria for septic shock and sepsis-associated acute kidney injury. 
We also discussed the research prospects of applying visual analytics to clinical trials.
}

\keywords{Visual Analytics, Healthcare, Clinical Trials, Decision Making, Electronic Health Record (EHR)}

\teaser{
  \centering
  \includegraphics[width=\linewidth]{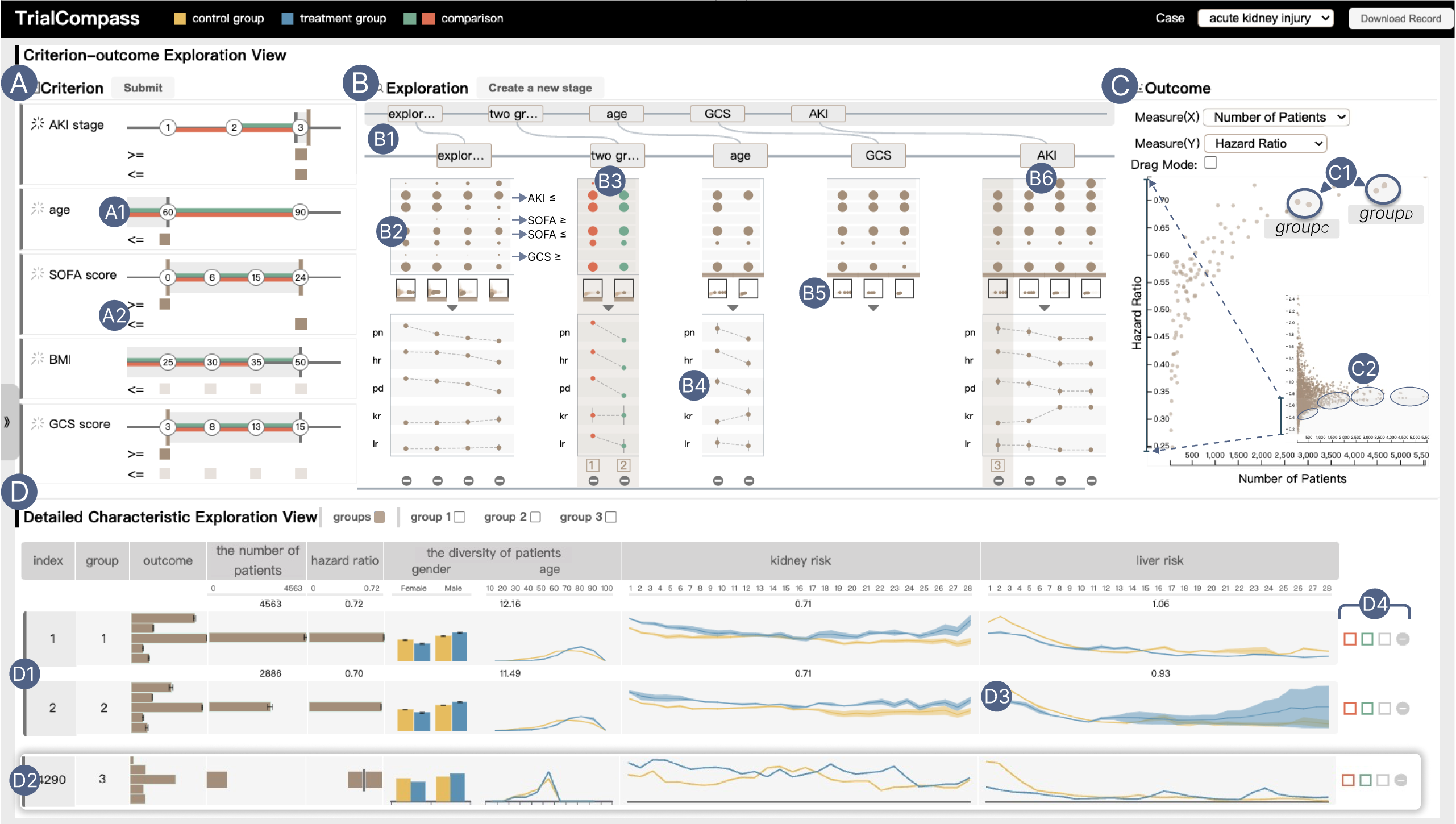}
  \caption{
    \textit{\systemname} enables clinicians to explore the vast space of eligibility criteria for clinical trials (approximately five thousand criterion candidates in this case).
    (A) Clinicians can adjust eligibility criteria based on their expertise or preferences in the Criterion View. (B) They can analyze the relationship between eligibility criteria and outcome metrics by creating stages throughout the exploration process, with history recorded for both. (C) Clinicians can explore the outcome metrics of all the criterion candidates in the Outcome View. (D) Clinicians can examine the detailed characteristics of the original EHR data for either individual candidates within a group or the summarization of a candidate group in the Detailed Characteristic Exploration View.
  }
  \label{fig: teaser}
  \vspace{0.1cm}
}

\graphicspath{{figs/}{figures/}{pictures/}{images/}{./}} % where to search for the images

\usepackage{tabu}                      % only used for the table example
\usepackage{booktabs}                  % only used for the table example
\usepackage{lipsum}                    % used to generate placeholder text
\usepackage{mwe}                       % used to generate placeholder figures
\usepackage{multirow}
\usepackage[normalem]{ulem}
\useunder{\uline}{\ul}{}

\usepackage[dvipsnames,svgnames]{xcolor}
\usepackage[most]{tcolorbox}
\usepackage{graphicx}
\usepackage{enumitem}
\usepackage{setspace}
\usepackage{float}
\setlist{noitemsep,parsep=0pt,partopsep=0pt} 
\newcommand{\eg}{e.g.}
\newcommand{\ie}{i.e.}

\newcommand{\systemname}{\textit{\textit{TrialCompass}}\xspace}

\newcommand{\Ea} {\textit{$E_{a}$}\xspace}
\newcommand{\Eb} {\textit{$E_{b}$}\xspace}
\newcommand{\Ec} {\textit{$E_{c}$}\xspace}
\newcommand{\Ed} {\textit{$E_{d}$}\xspace}
\newcommand{\Ee} {\textit{$E_{e}$}\xspace}
\newcommand{\Ef} {\textit{$E_{f}$}\xspace}
\newcommand{\Pa} {\textit{$P_{a}$}\xspace}
\newcommand{\Pb} {\textit{$P_{b}$}\xspace}
\newcommand{\Pc} {\textit{$P_{c}$}\xspace}
\newcommand{\Pd} {\textit{$P_{d}$}\xspace}
\newcommand{\Pe} {\textit{$P_{e}$}\xspace}

\newcommand{\review}[1]{\textcolor{purple}{}}
\usepackage{mathptmx}                  % use matching math font
\usepackage{CJK}

\newcommand{\xingbo}[1]{\textcolor{black}{#1}}
\newcommand{\revise}[1]{\textcolor{black}{#1}}

\definecolor{review}{HTML}{f72585}
\definecolor{tvcghighlight}{RGB}{18, 107, 174}
\definecolor{border}{RGB}{100, 100, 100}

\newtcbox{\mybox}[1][border]
  {on line, arc = 1pt, outer arc = 1pt, colframe = border,
    colback = #1!4!white, boxsep = 0pt, left = 1pt, right = 1pt, top = 1pt, bottom = 1pt, boxrule = 1pt}
    
\usepackage{marginnote}
\usepackage{adjustbox}
\global\marginparsep=9pt
\newcommand{\sidecomment}[1]{%
  \ifdefined\revise
  \marginnote{%
    \textcolor{review}{%
    }
  }
  \fi
}

\AtBeginDocument{

}

\begin{document}
\maketitle
\section{Introduction}
Clinical trials are studies on human subjects that assess the safety and effectiveness of new medical interventions (\eg, vaccines or drug usage), significantly advancing medicine. 
Completing a clinical trial is costly, often requiring around \$2.87 billion \cite{blass2015basic}.
Unfortunately, 86\% of clinical trials fail in the initial step due to the inability to recruit suitable participants within the specified timeframe \cite{huang2018clinical}. 
% This substantially impacts the progress of medical advancement. 
A major factor is the lack of well-designed \textbf{eligibility criteria}, including inclusion and exclusion criteria, to determine who is eligible to participate in the study \cite{liu2021evaluating}.
Designing effective eligibility criteria is a challenging task. 
Stricter criteria may hinder enrollment, while more relaxed ones can increase risks of adverse outcomes (\eg, worsening of the illness or even death) for target interventions.
Data-driven approaches \cite{liu2021evaluating, kwee2023target, kim2021towards, Fang2023data} have been proposed to help clinicians design more inclusive, safe, and effective criteria.
For example, Trial Pathfinder \cite{liu2021evaluating} can simulate the impact of adjusting a criterion (e.g., changing the age requirement from under 60 to under 70) on potential trial outcomes.
This information offers valuable insights for designing eligibility criteria.

Despite the availability of these tools, clinicians still encounter difficulties in designing suitable eligibility criteria.
\revise{First, existing tools cannot help clinicians efficiently \textbf{explore the large space of potential criterion combinations}—that is, the extensive set of possibilities generated by adjusting various eligibility criteria.
In practice, clinicians must assess how different combinations of criteria—such as age limits, medical history, and drug dosages—affect patient selection and outcomes. 
Each individual criterion often needs to be tested across multiple plausible settings based on clinical expertise.
For example, age may be set to under 60, 65, or 70 years, while medical history may specify no surgery in the past one, three, or six months.
This results in a combinatorial explosion of criterion candidates that are hard to explore.}
\sidecomment{R2C1, R2C2}
Second, existing tools fail to \textbf{provide contextual information on the temporal changes in patients' conditions} during the trial (e.g., whether liver risk consistently increases over time).
They only help clinicians trade off different aggregated metrics of trial outcomes, such as the number of patients that can be recruited versus the final trial hazard ratio. 
However, clinicians still need to consider finer details.
For instance, clinicians might weigh the final hazard ratio against changes in liver risk during the trial. 
Even with a low hazard ratio, they may still reject the eligibility criteria if they observe a persistent increase in liver risk.
Third, the lack of support to \textbf{track the exploration history} in eligibility criteria and outcome metrics throughout the iterative exploration process poses a significant challenge.
With numerous exploration operations, the iterative design process can quickly become highly complex.
Therefore, this tracking is crucial for iterative design processes. 
Without clear records of how changes in criteria impact trial outcomes, clinicians cannot systematically refine them and confidently determine the final settings based on the explored candidates.

To address the above challenges, we developed a visual analytics system named \systemname to assist clinicians in designing eligibility criteria. 
To the best of our knowledge, we are the first to leverage visualization techniques to address such a significant problem in the healthcare domain.
We first interviewed five experts to derive design requirements and developed a novel visual analytics workflow that enables clinicians to explore the expansive design space of eligibility criteria iteratively.
Given that experts may employ their expertise to varying degrees during the exploration, this workflow provides two approaches: knowledge-driven and outcome-driven.
The knowledge-driven approach enables experts to simulate outcomes by specifying different eligibility criteria based on their expertise. 
On the other hand, the outcome-driven approach supports experts in first examining a large number of criterion candidates to make an informed decision.
These two approaches offer clinicians flexibility in iteratively refining the eligibility criteria.
We have also integrated a history-tracking feature, allowing clinicians to trace their exploration process and understand the relationships between eligibility criteria, outcome metrics, and temporal details.
In summary, we made the following contributions:
\begin{itemize}[leftmargin=1em]
    \item We formulate the system design requirements for eligibility criteria design of clinical trials through collaboration with five experts in various specializations.
    \item We propose \systemname, a system integrating a novel workflow for iteratively exploring the large space of eligibility criteria through knowledge-driven and outcome-driven approaches. 
    \item We utilize a real-world dataset to conduct expert interviews and case studies, discovering novel insights for two important diseases (\ie, septic shock and sepsis-associated acute kidney injury).
\end{itemize}

\section{Related Work}
\subsection{Clinical trial design studies}
Clinical trials are crucial for assessing the safety and efficacy of new medical treatments. Designing suitable eligibility criteria is vital for trial success, impacting participant recruitment and final results \cite{Claessens2013Are}. 
These criteria specify the conditions that participants should meet to be eligible for a clinical trial, often including factors such as age, gender, medical history, and current health status.
However, designing criteria is a challenging task for clinicians since subtle adjustments in criteria may lead to big differences. 
For example, strict ones previously led to 80\% of advanced non-small-cell lung cancer patients being excluded from trials, significantly contributing to an 86\% deficit in meeting recruitment goals \cite{fehrenbacher2009randomized, huang2018clinical}. Researchers have thus explored data-driven approaches to support designing eligibility criteria more inclusively and effectively \cite{liu2021evaluating, kwee2023target}. 
Specifically, these approaches utilize historical patient data to measure the potential outcomes of specified criteria, such as the number of eligible participants for recruitment and their efficacy.
For example, Trial Pathfinder \cite{liu2021evaluating} measures the number of eligible patients and the hazard ratio of the setting criteria.
Moreover, it reveals a criterion's impact on a trial's outcomes.
\sidecomment{R2C3, R2C4}
\revise{However, current tools mainly rely on clinicians’ expertise to iteratively generate hypotheses, without fully leveraging precomputed outcome variations under different eligibility criteria settings. In this trial-and-error process, the lack of effective visual support often causes clinicians to lose contextual understanding and makes it difficult to assess how individual criteria or their combinations impact outcomes, hindering rigorous and informed decision-making.
Moreover, previous approaches fall short in supporting the comparison and balancing of multiple objectives (e.g., hazard ratios and patient counts). When conflicts arise, clinicians must manually make trade-offs, sometimes even consulting the original data, which is time-consuming.
Adding to the complexity, desirable outcomes often vary depending on clinicians’ goals. Their thinking may also evolve during the exploration process, making it difficult for automated algorithms to adapt effectively. These challenges highlight the need for human-in-the-loop tools that combine computational power with appropriate visualization techniques.
}

\subsection{Visual analytics for clinical data}
The deployment of visualizations in clinical research is becoming essential~\cite{sheng2025design, Sultanum2023ChartWalk, Linhares2023ClinicalPath, Debek2017timeline, Faiola2011Advancing, Hirsch2014HARVEST, Wang2022EHRSTAR, Zhang2019IDMVis}.
\sidecomment{R3C2}
\revise{For example, Wang et al.~\cite{Wang2022EHRSTAR} summarized visualization techniques for EHR data, which is one of the most significant and common data formats in the clinical domain.}
These visualization methods can help analyze the inherent complexity of clinical data and support critical decision-making in this high-stakes field.
\sidecomment{R1C5}
\revise{In this context, AI is increasingly integrated into clinical workflows, and visualization often serves as a bridge between AI and clinicians by addressing issues of uncertainty and interpretability~\cite{Cheng2022VBridge, Jiang2024HealthPrism, Kwon2019RetainVis}. For example, Cheng et al.~\cite{Cheng2022VBridge} proposed using visualization to help clinicians link original data with AI-generated features for improved decision-making. 
In addition to AI, traditional statistical models remain widely used and are often more thoroughly validated in clinical practice. However, visualization is still essential for helping clinicians understand the large volumes of data these models produce~\cite{Wang2022ThreadStates, Kwon2021DPVis, Kuo2023Animal}.
For example, DPVis~\cite{Kwon2021DPVis} employs Hidden Markov Models to calculate disease progression pathways, using visualization to provide clinicians with a more intuitive understanding of disease dynamics.}
Despite the evident benefits of visualization tools in enhancing various clinical workflows, their integration into clinical trial design remains limited. Although a recent study~\cite{li2023trialview} takes an initial step, it just visualizes patients' temporal progression during the treatment. It cannot be applied to complex decision-making scenarios like eligibility criteria design. The application of visualizations has great potential to facilitate exploring the huge design space of eligibility criteria.

\subsection{Visual analytics for multi-objective decision making}
Our work aims to assist clinicians in refining eligibility criteria considering multiple objectives, which is a multi-objective decision-making problem. 
In the visualization community, considerable visual analytic technologies have been explored to address multi-objective decision-making problems \cite{zhao2017skylens, weng2018homefinder, liu2017smartadp, chen2024fslens, Weng2019SRVis, Carenini2004ValueCharts, Wall2018Podium, Gratzl2013LineUp, Pajer2017WeightLifter, Wang2025SynthLens}. A key aspect of these tools is the ability to help users narrow down their choices from an immense selection of options. One of the strategies is to enable users to define their constraints and preferences. This approach allows users to see the effects of their constraints and filter the options accordingly \cite{weng2018homefinder, chen2024fslens}. 
However, it is challenging for experts to define constraints when designing eligibility criteria in a clinical trial context. 
Another strategy focuses on the discovery of user preferences through a process of heuristic exploration \cite{zhao2017skylens, Weng2019SRVis, Pajer2017WeightLifter, Gratzl2013LineUp}. 
This technique is particularly useful when users are unable to explicitly articulate their constraints or preferences. 
However, those works cannot support experts in systematically exploring the impact of various eligibility criteria on different outcome forms (\ie, multiple outcome metrics and original EHR data insights) and make informed decisions.
Therefore, it is urgent to develop a new visual analytics workflow that helps clinicians navigate through the large space of eligibility criteria.
\section{Design Study}
We developed a system to help clinicians design eligibility criteria. Over the past six months, we have collaborated closely with five domain experts (\Ea-\Ee) with various clinical specialties.
\Ea is a urologist with over twenty years of clinical trial experience. 
\Eb and \Ec are nephrologists with approximately three years of experience.
\Ed is a professor who has dedicated five years to data-driven clinical trial design.
Lastly, \Ee, a doctoral student who specializes in ophthalmology with two years of expertise.
We conducted one-hour semi-structured interviews with each expert to understand the eligibility criteria design challenges. From their requirements, we derived visual analysis tasks, then validated these through bi-weekly prototype feedback sessions. This study received IRB approval.

\subsection{Factors Related to Eligibility Criteria Design}
Data-driven design leverages historical patient EHR data to help define and refine eligibility criteria. For example, when testing a new drug, clinicians can explore past patient records to identify individuals who have taken similar compounds, using them as a reference for eligibility adjustments.
Specifically, clinicians often first determine \textit{eligibility criteria} to filter qualified patients based on their \textit{original EHR data} and assign \textit{medical interventions} to categorize them into \textit{treatment and control groups}. 
Then, several \textit{outcome metrics}, such as the hazard ratio and kidney risk ratio, can be calculated based on the original EHR data from the patients in the treatment and control groups. 
The \textit{temporal details} extracted from EHR data aid in interpreting these outcome metrics and understanding the more nuanced results of the intervention. 
These factors are crucial to the design of eligibility criteria, ensuring that its results are reliable, valid, and applicable to the intended patient population.
We have introduced those factors in detail as follows.

\textbf{1) Eligibility criteria} are the conditions designed by clinicians to recruit participants for clinical trials, categorized into inclusion and exclusion criteria \cite{liu2021evaluating}. 
Inclusion criteria define participant eligibility for a clinical trial, whereas exclusion criteria identify disqualifying traits.
Eligibility criteria may restrict participant demographics (\eg, age being between 18 and 70), health status (\eg, specific disease diagnoses, or recent medication history), and other relevant variables.
% Clinicians can determine criteria according to their domain knowledge and historically relevant clinical trials.
% However, this might introduce bias \cite{Kim2017BroadeningEC}.
% Data-driven approaches can enhance the design process of clinical trials.
% Despite these advancements, all the experts (\Ea-\Ee) indicated that determining suitable criteria remains difficult since they might have a range of uncertain criteria, each with numerous possible options. 
We define each unique combination of criteria with specific settings as a \textbf{criterion candidate} (e.g., age under 70, no heart surgery in the past three months, BMI under 30).

\textbf{2) Treatment and control groups} are differentiated by whether the enrolled participants receive the \textbf{medical intervention} being studied. The treatment group is given the medical intervention, while the control group receives a standard intervention (\eg, placebo) or no intervention \cite{Aggarwal2019StudyDP}. Then clinicians will compare these two groups to assess the effectiveness and safety of the medical intervention.

\textbf{3) Outcome metrics} assess the potential results in a clinical trial from systematic analysis of the treatment and control groups \cite{ferreira2017types}.
% Data-driven approaches enable clinicians to estimate interesting outcomes of a trial by leveraging historical patient datasets and their predefined eligibility criteria. 
Once clinicians establish their eligibility criteria, they can filter out qualified patients from historical patient datasets.
Then they identify a medical intervention that is either identical to or a suitable proxy for the current medical intervention being studied. 
This intervention allows them to categorize the filtered patients into treatment and control groups. 
Next, they measure various outcome metrics based on the two groups.
Through the literature survey, two outcome metrics (\ie, the number of patients and hazard ratio) are always seen as the primary focus of interest in clinical trial studies and calculated in data-driven approaches \cite{kwee2023target, liu2021evaluating}.
We have introduced them in detail as follows.
\begin{itemize}
    \item \textbf{The number of patients} refers to the total size of participants qualified for the clinical trial. Recruiting sufficient participants is critical to determine whether a clinical trial can proceed \cite{Desai2020recruitment}. All five experts underscored the importance of the number of patients.
    \item \textbf{Hazard ratio} refers to the ratio of hazard rates between the treatment and control groups, where a value less than one indicates a positive effect for the treatment group \cite{Royston2013Restricted}.
    This metric is often a direct indicator of the effectiveness of a medical intervention \cite{Opal2013Effect, Berk2020Effect, liu2021evaluating, kwee2023target}. 
    % For example, it can be measured through mortality rates of patients, such as 30-day mortality or 90-day mortality \cite{Chen2023Aspirin}.
    \sidecomment{R2C5}
    \revise{Additionally, it is typically reported with a p-value to assess statistical significance. Its clinical acceptability varies by context—for instance, in the treatment of rare diseases, even a hazard ratio slightly below 1 may be considered meaningful.}
\end{itemize}
We then conducted expert interviews to identify additional metrics of interest.
First, several experts suggested examining the overall diversity of the recruited participants.
Moreover, since most clinical trials focus on drug-based experiments, the experts were particularly concerned about the impact of drug usage on patients' organ function. 
They suggested evaluating kidney and liver function since the kidney is the primary organ responsible for the excretion of most drugs \cite{Ettore2016Adverse} and the liver also plays a crucial role in drug metabolism \cite{Anna2016Adverse}.
Specifically, they would like to understand the kidney and liver risks of patients over time.
% Therefore, we also involved two metrics related to kidney and liver risk to provide an average risk assessment over a period of time.
Additionally, the Charlson Comorbidity Index was mentioned to indicate patient mortality risk. However, due to a lack of relevant data on most patients, we decided not to incorporate this index. Below, we outline the additional outcome metrics used in our study.
\begin{itemize}
    \item \textbf{The diversity of patients} refers to the demographic breadth represented by participants who fulfill the eligibility criteria, encompassing a variety of attributes such as age, gender, race, and others. Recruiting a patient population with diverse demographic characteristics can lead to more broadly applicable results \cite{Kelsey2022Inclusion}. A greater value of diversity indicates a larger variety within the studied patient population.
    \item \textbf{Kidney risk ratio} and \textbf{liver risk ratio} measure the incidence of adverse events related to the kidney and liver between the treatment and control groups, respectively.
    A value less than one suggests that the treatment group has a lower risk of experiencing adverse events. \Ea, \Eb, and \Ec highlighted the necessity of evaluating the potential adverse reactions in survivors in the two groups. \Ea mentioned that the hazard ratio usually reflects the survival rate difference between the treatment and control groups, which is a primary concern of a clinical trial. However, it is also important to comprehend the health condition of those surviving patients. The kidney and liver are the two most critical concerns.
\end{itemize}
Clinicians must balance the five outcome metrics when defining eligibility criteria for clinical trials, as these metrics can at times present conflicting priorities. For instance, relaxing the eligibility criteria may increase participant enrollment, yet also result in a higher hazard ratio. As such, clinicians need to carefully weigh the tradeoffs among the five outcome metrics during the eligibility criteria design process.

\textbf{4) Temporal details} are derived from the original EHR data. 
The five outcome metrics are aggregation values calculated through patient cohorts. 
Therefore, clinicians also need to examine the temporal detailed characteristics in those patients' conditions to gain a more comprehensive understanding to compare different criterion candidates.
For example, \Ec highlighted the importance of tracking changes in kidney and liver function over time, as the medical intervention may have varying onsets across participants.

\begin{figure*}[ht]
\centering
\includegraphics[width=\linewidth]{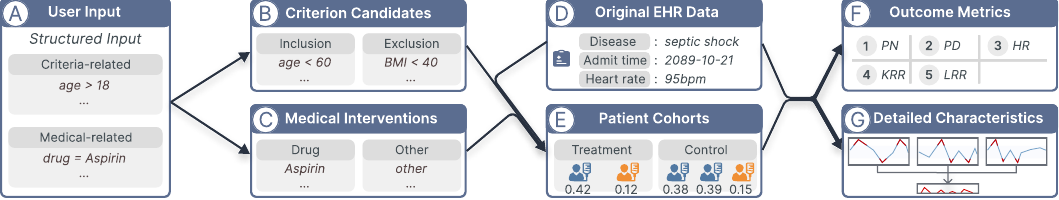}   
\vspace{-0.5cm}
\caption{(A) Clinicians input eligibility criteria and medical interventions. (B) Different criteria specifications can generate various criterion candidates. (C) The medical interventions will be used to divide patients into different groups. (D) The original EHR data of patients. (E) The system filters qualified patients and categorizes them into treatment and control groups based on eligibility criteria, medical interventions, and their original EHR data. (F) Measure the outcome metrics. (G) Organize the temporal detailed characteristics of the original EHR data.}
\label{fig: algorithm}
\vspace{-0.5cm}
\end{figure*}

\subsection{Visual Analysis Tasks}
\sidecomment{R1C1}
\revise{Based on our interviews with five experts, we have abstracted this domain problem into a multi-objective decision-making problem. 
Then, two authors performed independent inductive coding of interview transcripts using the thematic analysis methodology~\cite{Braun2006Thematic}. Through iterative coding and regular discussions, we reconciled interpretations and collectively distilled five key analytical tasks through consensus.}
\begin{itemize}[leftmargin=1.8em]
    \item[\textbf{T1}] \textbf{Support eligibility criterion specification for candidate generation.}
    We need to enable experts to make initial settings for eligibility criteria. Moreover, experts should be able to specify which criteria are adjustable based on their requirements and the range of possible adjustments for each criterion. Following their settings, we can generate a series of criterion candidates.
    \item[\textbf{T2}] \textbf{Present the outcomes of criterion candidates.}
    \revise{Some experts (\Ea, \Eb, and \Ee) emphasized the need to efficiently evaluate criteria configurations. As \Eb noted, \textit{``Manually testing different criteria combinations is prohibitively time-consuming.''} \Ea added, \textit{``Seeing potential outcomes of various criteria candidates in advance can help me form testable hypotheses.''}
    Therefore, our system needs to precompute and visually present all five outcome metrics for each candidate, enabling rapid comparative assessment.}
    \item[\textbf{T3}] \textbf{Support knowledge-driven and outcome-driven exploration.}
    \revise{All five experts emphasized the need to integrate their expertise when evaluating criterion candidates. For instance, \Ea explained, \textit{``For several criteria, I have prior expectations from literature or other clinical trials. The system should let me quickly validate these hypotheses.''}
    Experts also highlighted the benefits of using computed outcome metrics to address knowledge gaps. 
    \Ee mentioned that insufficient experience could be mitigated by leveraging big data to enhance exploration efficiency, a sentiment echoed by \Eb. 
    Consequently, our system should offer two exploration approaches: one that taps into clinicians' prior knowledge (knowledge-driven) and another that relies on measured outcomes (outcome-driven).}
    \item[\textbf{T4}] \textbf{Incorporate outcome metrics and temporal details for comparison.}
    \Ea, \Eb, and \Ec highlighted the need to examine details of the original EHR data. 
    \revise{For instance, \Ea mentioned, \textit{``We also need to track organ function trajectories—not just snapshot values. If kidney function declines steadily post-treatment, this signals intolerable toxicity even if the aggregate risk ratio appears acceptable.''}
    Therefore, it is crucial to facilitate the comparison of outcome metrics with temporal details.}
    \item[\textbf{T5}] \textbf{Facilitate the iterative navigation of criterion candidates.}
    After exploring outcome metrics and temporal details of original EHR data, experts need to gain insights that could lead to further refinement of eligibility criteria.
    \revise{Given the extensive exploration history generated, experts \Ea, \Ec, and \Ed emphasized the importance of systematically tracking and organizing this information.}

\end{itemize}{}

\section{Data Analysis}
\subsection{Data Description}
\sidecomment{R3C4}
\revise{In this work, we utilized the MIMIC-IV dataset~\cite{johnson2023mimic} as a historical record of patient data to study eligibility criteria design.}
The MIMIC-IV dataset is a publicly available database that provides comprehensive clinical data from intensive care units (ICUs). 
It also includes de-identified electronic health records (EHR) of patients. Specifically, it contains detailed information on over a hundred thousand patients, such as specific diagnoses, observed values of physiological indicators, and medication history. This information allows us to ensure a sufficient sample size when studying different clinical trials.

\subsection{Data Processing}
We first use the eligibility criteria and specific medical intervention entered by experts to separate patients into treatment and control groups (\autoref{fig: algorithm}-A-E).
In this process, adjustments to the criteria will generate multiple criterion candidates and result in different patient compositions of the two groups.
Then, our system can calculate the outcome metrics for each criterion candidate (\autoref{fig: algorithm}-F).
Finally, the temporal details of the treatment and control groups derived from the original EHR data will be systematically compiled and organized (\autoref{fig: algorithm}-G).

% \begin{figure*}[ht]
% \centering
% \includegraphics[width=\linewidth]{figs/algorithm.pdf}   
% \vspace{-0.5cm}
% \caption{Pending}
% \label{fig: algorithm}
% \vspace{-0.5cm}
% \end{figure*}

\subsubsection{Patient Cohort Construction}
First, we can identify eligible patients based on structured eligibility criteria entered by clinicians and classify them into treatment and control groups (\autoref{fig: algorithm}-E).
% Besides, we allowed clinicians to specify the eligibility criteria that require adjustment and the corresponding ranges for adjustment.
% This will generate a comprehensive list of criterion candidates.
When comparing the treatment and control groups, it is crucial to address the inherent differences in confounding factors, which are unrelated to the factors being studied. 
These confounding factors can introduce bias and affect the interpretation of the results.
Therefore, we performed a matching process for the two groups under each criterion candidate.
To achieve this, we leveraged the propensity score matching algorithm, like Trail Pathfinder \cite{liu2021evaluating}, to reduce bias caused by confounding factors (\eg, race, gender, and birthplace).
The propensity score is defined as the conditional probability of receiving the medical intervention given a set of observed confounding factors~\cite{rosenbaum1983propensity}.
\sidecomment{R3C5}
\revise{Specifically, this metric is used to identify matching patients within the control group to correspond with those in the treatment group, ensuring that their characteristics are comparable and allowing for a more accurate estimation of the treatment effect, which can be measured as follows}:

\vspace{-5mm}
\begin{align*}
e(F) &= P(T=1 \mid F),
\end{align*}
\vspace{-5mm}

\noindent where $e(F)$ represents the propensity score, $T$ denotes the medical intervention assignment (1 for treatment, 0 for control), and $F$ represents the confounding factors. 
Then, we iterated through each patient in the treatment group and identified the most similar patients in the control group based on their propensity scores (\autoref{fig: algorithm}-E). 
We then compared the propensity score difference between the matched pairs with a specified caliper value.
The caliper value, often set as the median absolute deviation of the propensity scores \cite{Austin2011Optimal}, serves as a threshold for acceptable similarity.
If the difference in propensity scores is not greater than the caliper value, the pair is considered a match.
Or we discard this particular sample.
Finally, these pairs will be used for subsequent comparison between the treatment and control groups.

\subsubsection{Outcome Metric Calculation}
We calculated five outcome metrics to assess the potential results of each criterion candidate (\autoref{fig: algorithm}-F).
\textbf{The number of patients} is the count of qualified patients. \textbf{The diversity of patients} is calculated based on gender and age, which are commonly mentioned in data-driven eligibility criteria design \cite{liu2021evaluating}. 
% It can also consider additional factors such as race based on experts' needs. 
We calculated the Shannon entropy based on the two attributes to represent the diversity of patients. 
We did not choose other diversity metrics, such as the Gini coefficient and Simpson index.
This is because these metrics have lower values when indicating higher diversity, which contradicts experts' intuition. 
% Therefore, we chose to use Shannon entropy as it is a measure of diversity that is more familiar to experts.

Then, we calculated the \textbf{hazard ratio} based on the treatment and control groups through training the Cox proportional-hazards model \cite{spotswood2004hazard}, which is a commonly used method for survival analysis. Specifically, the model is formulated as follows:

\vspace{-5mm}
\begin{align*}
h(t|X) = h_0(t) \cdot e^{\left(\beta_1X_1 + \beta_2X_2 + \ldots + \beta_pX_p + \beta_TT\right)},
\end{align*}
\vspace{-5mm}

\noindent where $h(t|X)$ represents the hazard function at time $t$ given the covariates $X$, $h_0(t)$ represents the baseline hazard function, and $\beta_1, \beta_2, \ldots, \beta_p$ correspond to the regression coefficients associated with each covariate. Moreover, $T$ indicates the medical intervention assignment. Finally, the hazard ratio can be represented by $HR = e^{(\beta_T)}$.

For the \textbf{kidney risk ratio and liver risk ratio}, we used serum creatinine (SCr) \cite{Bostom2002Predictive} and aspartate aminotransferase (AST) \cite{Pratt2000Evaluation} as indicators, respectively. 
We extracted them from the original EHR dataset.
Due to potential truncation, some patients may have missing or incomplete data. We addressed this by imputing values based on discharge status: for discharged patients, we replaced missing data with normal values, assuming recovery.
For patients with missing data before discharge or death, we discarded samples with significant gaps and applied interpolation for the rest \cite{Dziura2013Strategies}. We then calculated daily kidney and liver risk ratios, averaging them to provide overall assessments for each.

\subsubsection{Temporal Detail Organization}
% In this section, we describe how we organize the raw data. 
Our experts (\Ea-\Ee) pointed out that it is challenging to understand the demographic characteristics of the recruited patients only through the diversity score. 
Therefore, we compiled the distribution of gender and age for both the treatment and control groups. 
In addition, the hazard ratio needs to be accompanied by a confidence interval to allow clinicians to understand the significance of the results. 
Furthermore, considering that the kidney and liver risk ratios represent an average of the risk over a period, clinicians need to examine how these risks evolve over time in the two groups.
Therefore, we summarized the risk degree for the organs over time (\autoref{fig: algorithm}-G).
For patients who were alive at a specific time, we calculated the average degree of abnormality in their indicators—such as SCr for kidney risk and AST for liver risk—for both the treatment and control groups, based on reference ranges.

% \vspace{-2mm}
% \[
% D(T, t) = \frac{1}{N_T(t)} \sum_{i=1}^{N_T(t)} d_i(t) \cdot \mathbbm{1}_{\{T_i = T\}},
% \]

% \noindent where $T$ indicates which group each patient belongs to and $t$ represents the time. $N_T(t)$ represents the number of living patients in the treatment or control group at time $t$. $d(t)$ represents the abnormality degree of a patient at time $t$, which can be calculated by:

% \vspace{-2mm}
% \[
% d(t) =  \begin{cases}
% v(t) - {I_{max}}, & \text{if } v(t) > {I_{max}}, \\
% 0, & \text{if } {I_{min}} \leq v(t) \leq {I_{max}}, \\
% {I_{min}} - v(t), & \text{if } v(t) < {I_{min}},
% \end{cases}
% \]
% \vspace{-2mm}

% \noindent where $I_{max}$ and $I_{min}$ represent the normal range of a certain indicator, such as SCr for the kidney risk or AST for the liver risk, and $v(t)$ denotes the value of this indicator at time $t$. This information can allow clinicians to realize the extent of deviation from normal values of the two groups at different times.

\section{Visual Design}
\systemname provides three views to support clinicians in designing eligibility criteria: the Criterion Specification View (\autoref{fig: criterion_specification_view}), the Criterion-outcome Exploration View (\autoref{fig: teaser}-A, B, C), and the Detailed Characteristic Exploration View (\autoref{fig: teaser}-D). 
% The Criterion Specification View allows clinicians to input eligibility criteria and medical interventions in a structured way. 
% Moreover, clinicians can specify which of these eligibility criteria require adjustments through designated ranges or categories.
% The other two views can help explore numerous criterion candidates with their outcome metrics and detailed characteristics.
% The Criterion-outcome Exploration View is divided into three sub-views: the Criterion View, the Outcome View, and the Exploration View.
% Clinicians can examine various criterion candidates based on their prior knowledge through the Criterion View (\ie, knowledge-driven) or based on the calculated outcome metrics through the Outcome View (\ie, outcome-driven). 
% Furthermore, the Detailed Characteristic View enables clinicians to examine detailed information about patients.

% \begin{figure}[ht]
% \centering
% \includegraphics[width=\linewidth]{figs/workflow.pdf}   
% \vspace{-0.5cm}
% \caption{Pending...}
% \label{fig: workflow}
% \vspace{-0.5cm}
% \end{figure}

\subsection{Criterion Specification View}
\begin{figure}[ht]
\centering
\includegraphics[width=\linewidth]{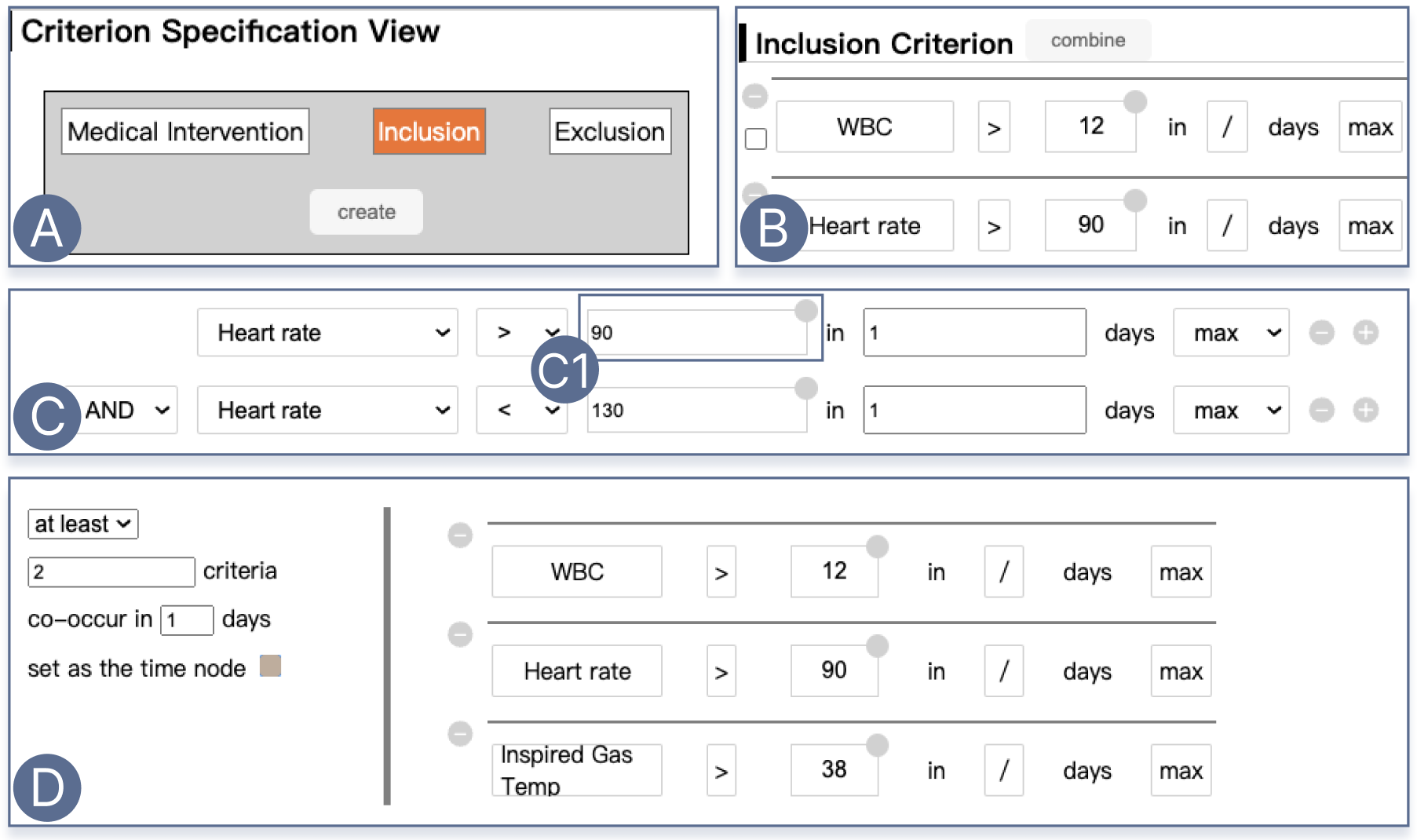}   
\vspace{-0.5cm}
\caption{(A-D) The Criterion Specification View enables clinicians to specify inclusion criteria, exclusion criteria, and medical interventions.}
\label{fig: criterion_specification_view}
\vspace{-0.5cm}
\end{figure}

This view facilitates clinicians in entering eligibility criteria for enrollment and defining the particular medical intervention that differentiates between treatment and control groups (\textbf{T1}). 
Clinicians can choose to create criteria for medical interventions, inclusion criteria, and exclusion criteria (\autoref{fig: criterion_specification_view}-A, B).
The system supports the use of ``AND'' and ``OR'' within a single criterion, as well as several aggregation functions like the minimum and maximum (\autoref{fig: criterion_specification_view}-C).
It also supports the combination of multiple criteria, such as ``at least two eligibility criteria must be met'' (\autoref{fig: criterion_specification_view}-D).
In addition, our system provides clinicians with user-friendly prompts, like displaying detailed explanations when hovering over an entity in a drop-down list.
Finally, our system allows clinicians to customize uncertain criteria that they would like to adjust.
They can click the corresponding button to set multiple adjustment values (\autoref{fig: criterion_specification_view}- C1). 
By combining these adjustments for each criterion, our system can generate criterion candidates, evaluate their outcome metrics, and organize the temporal details of the original EHR data.

\subsection{Criterion-outcome Exploration View}
This view allows clinicians to systematically navigate through a comprehensive array of criterion candidates along with their associated outcome metrics.
There are three sub-views: the Criterion View, the Outcome View, and the Exploration View.

\subsubsection{Criterion View}
This view (\autoref{fig: teaser}-A) enables experts to adjust each eligibility criterion through sliders based on their prior knowledge or insights during the exploration process (\textbf{T3}).

\textbf{\textit{Description:}} This view is organized sequentially, listing each eligibility criterion that experts need to adjust. 
A slider is provided for each criterion to allow clinicians to make adjustments.
If clinicians initially specify a maximum and minimum constraint for a criterion, two sliders will appear to represent these limits.
Additionally, to aid experts in understanding the range, the region of the effective range is highlighted.
Beneath each tick mark on the slider, color-coded visualizations indicate how many criterion candidates satisfy the corresponding constraint (\autoref{fig: teaser}-A2). 
They can interact with the scatter plot in the Outcome View to show the distribution of the selected candidates.
Considering that clinicians often need to compare two candidates, two colors are used above and below the slider to display the distinctions between the criteria of two candidates selected in the other views (\autoref{fig: teaser}-A1).

\textbf{\textit{Justification:}} Originally, we utilized a set of polylines along parallel sliders to display the combinations of various eligibility criteria (\autoref{fig: alternative_design}-A). 
However, this approach introduced visual confusion. 
Interleaved polylines made it difficult for experts to discern differences between two candidates. Furthermore, in this situation, the upper and lower limit requirements of a criterion will be separately displayed, which does not align with the customary practices of experts.

\subsubsection{Outcome View}
This view (\autoref{fig: teaser}-C) reveals the values of outcome metrics associated with all potential criterion candidates using a scatter plot (\textbf{T2}). 
Experts can select any two of the five metrics to represent on the axes. 
We avoided using glyphs to prevent visual clutter and potential interpretation difficulties. Instead, experts can set two axes to represent each criterion candidate as a point in the scatter plot, allowing for zooming during exploration. 
They can also lasso interesting candidates, which will be shown in the Exploration View for further analysis (\textbf{T3}).

\subsubsection{Exploration View}
This view (\autoref{fig: teaser}-B) enables clinicians to conduct systematic exploration. It allows experts to navigate through the vast space of eligibility criteria based on stages. In each stage, the exploration process is recorded through a snapshot, assisting clinicians in tracking and analyzing their exploration path (\textbf{T5}). This feature helps them to understand and narrow down the complex criterion space effectively.

\textbf{\textit{Description:}} This view helps experts understand the relationship between criteria and outcomes through their exploration history. Given the complexity of adjusting criteria in the Criteria View and selecting candidates in the Outcome View, systematically recording this operation history is essential for clarity.
To achieve this, we introduced stages (\autoref{fig: teaser}-B1). Experts can create a new stage whenever they consider an exploration action to be independent of previous ones. 
Their exploration will be recorded through a snapshot, where they can assign an importance level, keywords, and detailed descriptions to this stage.
This enables precise documentation and easy review of previous operations.
Furthermore, we provide a condensed stage visualization that displays all important stages at the top of the Exploration View. 

In each stage snapshot, our system visualizes the exploration history from two main operations: adjusting eligibility criteria in the Criterion View and selecting criterion candidates based on outcome metrics in the Outcome View. 
First, a matrix displays changes in eligibility criteria (\autoref{fig: teaser}-B2). Typically, each row in a matrix represents a criterion. When a criterion has a minimum and maximum threshold set, two rows will represent them respectively. 
Columns correspond to exploratory records, and circles indicate values—larger circles represent higher values, with exact figures displayed on hover.
Second, we present changes from selecting criterion candidates in the Outcome View using two methods.
Thumbnails of the scatter plot provide an intuitive overview of operations (\autoref{fig: teaser}-B5).
Considering that the scatter plot in the Outcome View might involve changing the horizontal and vertical axes, we display these axes in the thumbnail if they have been changed.
Additionally, line charts illustrate the average values of five outcome metrics during exploration (\autoref{fig: teaser}-B4), helping experts comprehend how different outcome metrics fluctuate throughout their operations.
% Finally, given that experts need to examine all the aforementioned information together for further exploration, our system will display all the visualizations for the two types of operations. 
% Specifically, when experts select criterion candidates in the Outcome View, the matrix will showcase the average value of each criterion.

\begin{figure}[ht]
\centering
\includegraphics[width=\linewidth]{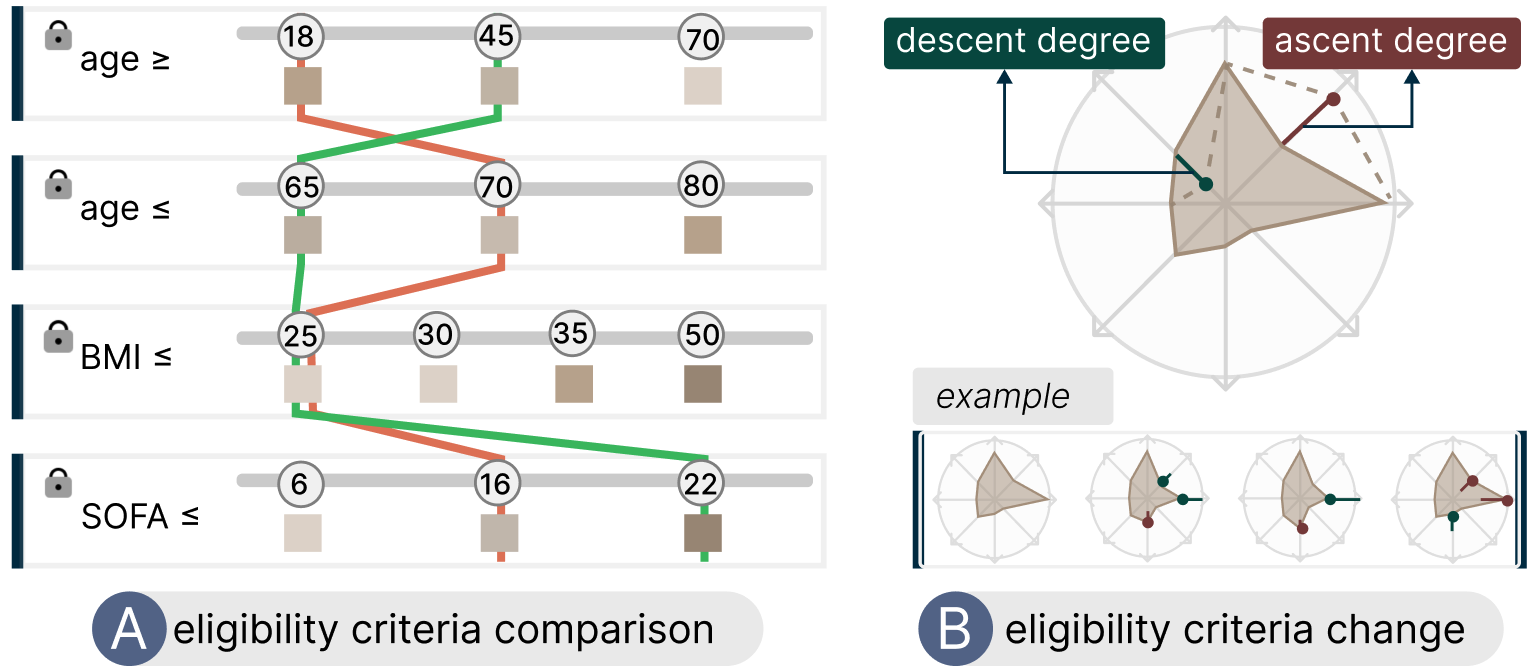}   
\vspace{-0.5cm}
\caption{(A) The alternative design is to compare the eligibility criteria of two individual criterion candidates. (B) The alternative design is to track the changes in multiple eligibility criteria.}
\label{fig: alternative_design}
\vspace{-0.2cm}
\end{figure}

\textbf{\textit{Justification:}} First, we initially leveraged line charts to display the change in eligibility criteria.
However, as the number of eligibility criteria increases, each line chart occupies a smaller portion of the space. 
This diminishes the visual amplitude of variations, making it challenging to discern changes.
Then, we used radar charts to represent changes in criteria (\autoref{fig: alternative_design}-B).
However, we discovered that experts encountered difficulty in comparing radar charts to understand the changes in a specific criterion over time.
Second, we just recorded how eligibility criteria changed and the five outcome metrics evolved over time with each operation conducted by clinicians at first. 
However, when faced with a large amount of historical data, experts often struggle to keep track of their previous operations and relevant records during the current exploration.
Therefore, we introduced the concept of stages. 

% \begin{figure}[ht]
% \centering
% \includegraphics[width=\linewidth]{figs/exploration_alternative_design.pdf}   
% \vspace{-0.5cm}
% \caption{Alternative design}
% \label{fig: exploration_alternative_design}
% \vspace{-0.5cm}
% \end{figure}

\subsection{Detailed Characteristic Exploration View}
This view (\autoref{fig: teaser}-D) presents temporal detailed characteristics of the original EHR data in two modes: group and individual (\textbf{T4}).
For the group mode, the visualization presents the average and standard deviation of temporal detailed characteristics for all the criterion candidates in a group (\autoref{fig: teaser}-D1).
The individual mode displays each criterion candidate in a group individually (\autoref{fig: teaser}-D2).
Specifically, it first displays the index and five outcome metrics. 
Additionally, histograms indicate the number of patients and hazard ratios. 
Different colored line charts represent the treatment and control groups for the distributions of gender, age, kidney function over time, and liver risk over time. 
Line charts are used instead of histograms to reduce clutter and enhance trend analysis.
Experts can select two groups in group mode or two candidates in individual mode for comparison (\autoref{fig: teaser}-D4). 
The criteria for the two selected groups are showcased in the Exploration View (\autoref{fig: teaser}-B3), while the criteria for two individual candidates are displayed in the Criterion View (\autoref{fig: teaser}-A1).

% \subsection{Interactions}
% To support clinicians complete the exploration, we have developed numerous interaction approaches in our visual analytics system.

% \textit{Selecting criterion candidates:}
% Our system provides two operations to support experts in selecting criterion candidates. The first is adjusting sliders in the Criterion View, and the second is using a lasso tool to select points within the scatter plot in the Outcome View.

% \textit{Creating a stage:} Our system allows experts to create a stage in the Exploration View to record the operation history independently. Besides, experts can specify the importance level, keywords, and details to a stage to assist their recall and understanding.

% \textit{Checking the detail characteristics:} Our system enables experts to examine the detailed characteristics of criterion candidates by clicking on records in the Exploration View. All selected candidates will then be displayed in the Detail Exploration View.

% \textit{Comparing detail characteristics:}
% Our system supports experts in dragging to reorder criterion candidates while viewing detailed characteristics in the Detail Exploration View. This can facilitate a quick comparison between two candidates.

% \textit{Recording the preference of criterion candidates:}
% When experts examine detailed characteristics in the Exploration View, our system allows them to mark criterion candidates they consider promising.

\section{Case Study}
\sidecomment{R3C4}
We have conducted case studies \revise{based on the MIMIC-IV dataset~\cite{johnson2023mimic}} for two different diseases (\ie, septic shock and sepsis-associated acute kidney injury).
\sidecomment{R1C4}
\revise{For the first case, we invited a new clinician \Ef, who has over five years of experience in clinical trials in sepsis.
For the second case, we invited our previous expert \Eb, who specializes in kidney disease research and has three years of experience in clinical trials.}
We thoroughly documented the exploration process of both experts to showcase how our system assists in eligibility criteria design.

\begin{figure*}[ht]
\centering
\includegraphics[width=\linewidth]{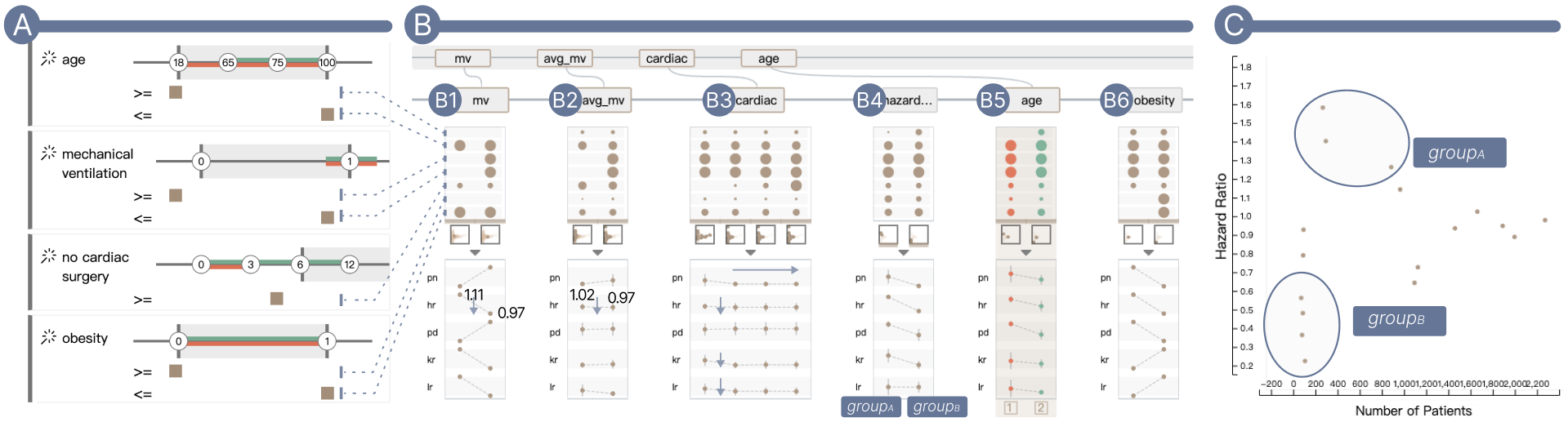}   
\vspace{-0.5cm}
\caption{(A) Specify four eligibility criteria. (B) The exploration process. (B1-B2) Investigate the impact of mechanical ventilation. (B3) Study the impact of cardiac surgery. (B4) Identify the factors (\ie, age and obesity) that can impact the hazard ratio through the outcome-driven approach.
(B5) Validate the impact of age.
(B6) Validate the impact of obesity. (C) Explore two groups of candidates with hazard ratios greater than or less than 1.}
\label{fig: case1_1}
\vspace{-0.5cm}
\end{figure*}

\subsection{Case I: Septic Shock}\label{case_one}
\sidecomment{R2C9}
\revise{\Ef is an expert in sepsis research and is very interested in a historical clinical trial\footnote{https://clinicaltrials.gov/study/NCT01448109} investigating the efficacy of \textit{\textbf{hydrocortisone}} (a kind of drug) in patients with septic shock. 
She desired to assess whether the eligibility criteria in this clinical trial could be refined. In addition, she would like to check whether and how to add two more criteria that are commonly seen in other related clinical trials.}

\textbf{Specifying the eligibility criteria (T1).} 
\revise{First, \Ef established the eligibility criteria and medical intervention based on this clinical trial.}
Among these, she focused on adjusting two specific criteria. 
She thought one was slightly relaxed, while another was too restrictive to potentially exclude some patients who could benefit. 
% 第一个条件
The first is \textbf{age} (the first row in \autoref{fig: case1_1}-A), as this clinical trial requires patients above 18 years old. However, she desired to understand how age might impact the efficacy of hydrocortisone. 
% \rui{For instance, she was considering whether this drug might not be as effective for older adults compared to younger individuals. This could influence her to set an upper age limit.} 
For instance, she was considering whether hydrocortisone might be less effective in older patients. This consideration could lead her to establish an upper age limit for the trial.
% 第二个条件
Secondly, she noticed that it specified the requirement for patients to be on \textbf{mechanical ventilation} (the second row in \autoref{fig: case1_1}-A), which usually indicates a severe condition. She wondered if hydrocortisone was also suitable for patients with less severe conditions.

% 第三\四个条件
Additionally, \Ef introduced two new criteria for adjustment. 
These criteria were not considered in the original trial but are frequently included in other related clinical trials. 
\Ef deemed them important based on her expertise.
The first was that patients should not have undergone \textbf{cardiac surgery} (the third row in \autoref{fig: case1_1}-A) within the past six months. 
From her expertise, patients who had previously undergone cardiac surgery and subsequently developed sepsis were at a higher risk of developing septic cardiomyopathy, which carried a higher mortality rate. 
Therefore, she hypothesized that hydrocortisone might not be as effective for these patients. 
Finally, she desired to assess whether to recruit patients with obesity, as \textbf{obesity} (the fourth row in \autoref{fig: case1_1}-A) can exacerbate organ damage caused by septic shock. 
She set the range of those criteria within the Criterion Specification View.
Initially, she set the criteria to match the original clinical trial, finding a hazard ratio of 1.225 with a statistically significant confidence interval. The kidney and risk ratios were around 2, indicating high stakes. Despite the large sample size, these metrics led her to refine the eligibility criteria.

\textbf{Knowledge-driven exploration (T3).} 
Initially, \Ef desired to examine whether patients who did not require mechanical ventilation could still benefit from hydrocortisone.
Therefore, she explored the impact of the mechanical ventilation requirement by creating a stage in the Exploration View and specifying the slider in the Criterion View. 
% 创建一个stage并且take notes
She found that the group without receiving mechanical ventilation (the first column in \autoref{fig: case1_1}-B1) showed a deterioration in various outcome metrics.
This was indicated by a higher hazard ratio, as well as higher kidney and liver risk ratios (the second, fourth, and fifth line charts in \autoref{fig: case1_1}-B1).
Upon hovering over the hazard ratio line chart, she noticed that this group particularly had a hazard ratio greater than 1 (1.11).
Conversely, the group that received mechanical ventilation (the second column in \autoref{fig: case1_1}-B1) had a hazard ratio below 1 (0.97).
\textit{``A hazard ratio below 1 indicates a positive effect of the treatment. Therefore, it suggests that hydrocortisone may not have a positive effect on patients without mechanical ventilation.''}
Then, \Ef would like to examine the potential outcomes from interactions between the mechanical ventilation criterion and other criteria, as she was concerned this might lead to changes in the effects.
Therefore, \Ef decided to analyze all the criterion candidates requiring mechanical ventilation alongside those not requiring mechanical ventilation. 
She found that the average hazard ratio of the former was less than 1 (the second column in \autoref{fig: case1_1}-B2, the second line chart). 
This reinforced her belief in following the historical clinical trial and targeting patients undergoing mechanical ventilation.
\textit{``Mechanically ventilated patients often have more severe systemic inflammation. In this context, hydrocortisone seems to be effective at modulating the inflammation and improving their symptoms.''}

\Ef proceeded to evaluate the requirement of duration after cardiac surgery.
She hoped to determine whether the implementation of this criterion could mitigate the hazard ratio and other risk ratios.
% 创建一个stage并且take notes
Therefore, she created a new stage to analyze the criterion candidates with varying time requirements after cardiac surgery. 
Her comparisons of all five outcomes across different time requirements (indicated in the line charts in \autoref{fig: case1_1}-B3) revealed that candidates without a time requirement (the first column in \autoref{fig: case1_1}-B3) had a higher average hazard ratio (greater than 1), kidney risk ratio, and liver risk ratio.
\sidecomment{R2C7}
\revise{This aligned with her prior knowledge that patients with a history of cardiac surgery who later developed sepsis faced an elevated risk of septic cardiomyopathy, which in turn carried a higher mortality rate.}
Furthermore, she noticed that there was almost no difference in the outcome metrics between the time requirements of 3, 6, or 12 months (the last three columns in \autoref{fig: case1_1}-B3, the line charts).
Therefore, \Ef decided to set the inclusion criterion requiring a duration after cardiac surgery greater than 6 months.
This decision was predicated on the medical understanding that a 3-month threshold typically denotes potential safety, whereas a 6-month threshold signifies a more fundamental level of safety. 
By implementing the 6-month benchmark, \Ef could ensure a baseline level of safety while still avoiding the exclusion of patients who could potentially benefit.

\begin{figure}[ht]
\centering
\includegraphics[width=\linewidth]{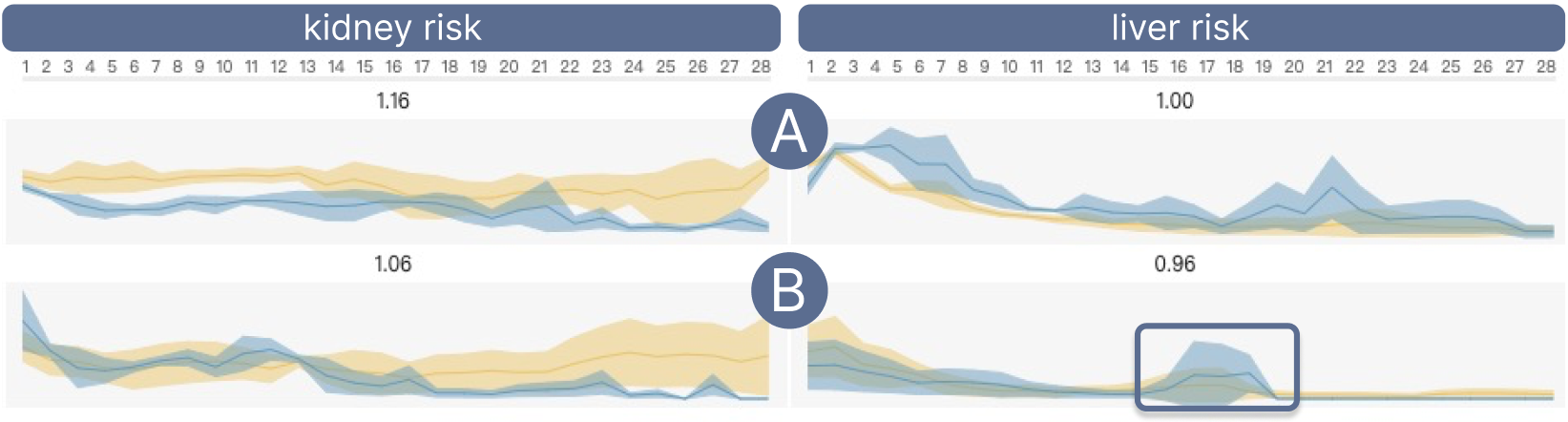}   
\vspace{-0.5cm}
\caption{The temporal details of the two groups. (A) The group with minimum age limits set at 18. (B) The group with minimum age limits set at 65. \Ef found that older patients exhibit lower kidney and liver risks, but caution is advised regarding the mid-term increase in liver risk.}
\label{fig: case1_2}
\vspace{-0.25cm}
\end{figure}

\textbf{Outcome-driven exploration (T3).}
After finalizing the two criteria, \Ef used the Outcome View to examine the factors influencing the hazard ratio. 
To understand how age and obesity jointly influence the outcome, she performed a lasso selection on the scatter plot based on y-axis values (\textbf{T2}) to compare two distinct groups: \texttt{$group_A$} (hazard ratio > 1) and \texttt{$group_B$} (hazard ratio < 1) (\autoref{fig: case1_1}-C1).
She discovered that the criterion candidates in \texttt{$group_B$} had a higher minimum age requirement and a greater number of patients with obesity (as indicated by the circle size in the first row and the last row of the matrix in \autoref{fig: case1_1}-B4).
She also checked it through the heatmap under each slider in the Criterion View.
\textit{``This indicates that recruiting patients with older age may lead to better treatment outcomes. Furthermore, patients with obesity can also benefit from the treatment.''}

Employing a similar approach as before, \Ef discovered that patients over the age of 65 (the second column in \autoref{fig: case1_1}-B5, the line charts) had better treatment efficacy, indicated by lower hazard ratios (\textbf{T5}).
She also observed that increasing age had a minimal impact on kidney and liver risk ratios (\autoref{fig: case1_1}-B5, the fourth and fifth line charts).
\textit{``Although there was minimal change in the kidney or liver risk ratios, it is still important to identify when these patients might experience abnormalities.''}
Therefore, \Ef delved into the Detailed Characteristic Exploration View (\textbf{T4}). 
She discovered that older patients in the treatment group were more likely to experience mid-term liver issues (as indicated by the steep increase in the blue line during the mid-term period in \autoref{fig: case1_2}-B). 
\textit{``This may be due to liver-related side effects that arise after administering a certain amount of hydrocortisone. I find it acceptable since these patients are expected to recover in the later stages.''} 
Finally, \Ef made an intriguing observation during her analysis of obesity: patients with obesity (the second column in \autoref{fig: case1_1}-B6) indeed displayed a favorable response to the treatment since their hazard ratio was lower. 
This led her to consider the presence of the obesity paradox. \textit{``Despite the potential adverse effects of obesity on organ function, several studies have indicated that individuals with obesity exhibit lower mortality rates. This phenomenon has been observed in some diagnosis scenarios. It seemed to be also present in this case.''}
However, she also emphasized the importance of monitoring the proportion of abnormalities in liver and kidney function, as indicated by the last two line charts in \autoref{fig: case1_1}-B6. Despite this, she ultimately decided to include patients with obesity, as the hazard ratio was remarkably low.

\textbf{The final decision.}
\sidecomment{R3C6}
\revise{Finally, \Ef reviewed all of her explorations again.} Using the stage-based visualization, she systematically rechecked the reason behind each decision and then summarized the insights.
She determined the following key criteria based on her findings: recruiting patients undergoing mechanical ventilation, setting a minimum time frame of 6 months after cardiac surgery, including patients with obesity, and enrolling as many elderly patients as possible. 
\textit{``This system has helped me improve potential outcomes compared to the historical eligibility criteria. My key concern—the hazard ratio—has shifted from greater than 1 to less than 1, which is very promising.''}
She also noted that the kidney and liver risk ratios are below 1, boosting her confidence in her decision-making
\textit{``Although patient recruitment has decreased, this likely filters out those who do not respond well to treatment.''}

\subsection{Case II: Sepsis-associated Acute Kidney Injury}
\label{case_two}
\Eb, an experienced nephrologist, was interested in studying the effects of aspirin on sepsis-associated acute kidney injury. He hoped to leverage our system to optimize five eligibility criteria.

\textbf{Specifying the eligibility criteria (T1).} He set the inclusion criteria as patients with sepsis-associated acute kidney injury. He then used aspirin to divide the treatment and control groups. 
Next, he selected five criteria which were always considered in kidney-related diseases and he was interested in (\autoref{fig: teaser}-A).
% 第一个
The first was the \textbf{AKI stage}, categorized into three levels, indicating the severity of kidney dysfunction.
% 第二个
The second was \textbf{age}. In some clinical trials, older patients are often excluded due to potential organ decline and reduced treatment compliance.
% 第三个
The third was the \textbf{SOFA score}, reflecting the degree of organ failure and providing insights into the patient's current health condition.
% 第四个
The fourth was \textbf{BMI}, which is used to evaluate whether an individual is within the healthy weight range. Excessively overweight patients are sometimes excluded from clinical trials for safety reasons.
% 第五个
The fifth was the \textbf{GCS score}, which is widely used in emergency medicine to assess a patient's level of consciousness.
% 初始结果
Then, \Eb inputted these criteria and manually corrected them.
Initially, the AKI stage was required to be over 1, and the age was limited to below 60. The SOFA score had to be less than 15, and no specific requirements were set for the GCS score. Additionally, patients whose BMI was larger than 35 were excluded from the study.
% \rui{These criteria resulted in approximately 1000 qualified patients.
% In addition, the hazard ratio was 0.59, with a confidence interval that did not include 1, indicating statistical significance.} 
\xingbo{These criteria identified approximately 1,000 eligible patients. The analysis yielded a statistically significant hazard ratio of 0.59. Additionally, the kidney risk ratio was below 1, while the liver risk ratio exceeded 1.}
% Besides, the kidney risk ratio was below 1, while the liver risk ratio was over 1.
\xingbo{\Eb considered the hazard ratio favorable but deemed the patient sample size insufficient. He hoped to greatly increase the patient enrollment, while still maintaining the low hazard ratio and other favorable outcome metrics.}

\textbf{Outcome-driven exploration (T3).}
Given the numerous criteria and their interactions, \Eb found it difficult to adjust the criteria and examine the outcomes. However, with so many potential candidates, selecting the ones for further examination was also challenging. 
Therefore, he decided to explore the relationship between the criteria and outcome metrics first to see how to reduce the exploration space.
Therefore, \Eb initially identified four regions (\autoref{fig: teaser}-C2) on the edges of the scatter plot where the hazard ratio and the number of patients were balanced (\textbf{T2}). \textit{``These regions represent a trade-off, where increasing the number of patients can lead to a higher hazard ratio.''}
From the records in the Exploration View, he observed that the size of the circles remained relatively unchanged in the second, fourth, fifth, and seventh rows of the matrix in \autoref{fig: teaser}-B2, respectively.
This indicated that the upper limit for the AKI stage, the lower and upper limit for the SOFA score, and the lower limit for the GCS score were stable in these four regions.
Consequently, he established these criteria (\ie, $the~AKI~stage\leq3$, $0\leq \textit{the SOFA score}\leq24$, and $the~GCS~score\geq3$) since he believed that within this range, he was more likely to find the criterion candidate he was satisfied with.
\Eb mentioned, \textit{``Patients with higher AKI stages indicate more severe kidney injury, while lower GCS scores suggest more significant brain impairment. This suggests that severely affected patients may benefit from aspirin. Furthermore, the overall impact of the SOFA score seems minimal.''}

Next, \Eb selected two smaller groups (\ie, \texttt{$group_C$} and \texttt{$group_D$}) for in-depth analysis (\autoref{fig: teaser}-C1).
Therefore, he examined their details in the Detailed Characteristic Exploration View (\autoref{fig: teaser}-D) (\textbf{T4}).
While $group_C$ exhibited a slightly lower hazard ratio (0.70) compared to $group_D$ (0.72), he noticed that the number of patients decreased by almost half (\autoref{fig: teaser}-D1).
Additionally, he noted that the treatment group in $group_C$ exhibited a continuous deterioration of liver function by the end of the experiment (\autoref{fig: teaser}-D3).
Therefore, he was inclined to choose $group_D$ for further exploration.
Then, \Eb examined the two individual criterion candidates in $group_D$.
Based on the comparison in the Criterion View, he found that both selected candidates had an upper age limit of 90 and an upper GCS score limit of 15 (\autoref{fig: teaser}-A). 
To validate this, \Eb adjusted the corresponding slider and discovered that there were not sufficient patients aged under 60 (as indicated by the first line chart in \autoref{fig: teaser}-B4) (\textbf{T5}). He further confirmed that setting the upper age limit to 90 was a preferable choice.
Using the same approach, \Eb determined the upper limit of the GCS score.

\textbf{Knowledge-driven exploration (T3).}
\Eb started to determine the remaining two eligibility criteria: the lower limit of the AKI stage and whether to enroll patients with high BMI. 
\Eb believed that patients with an AKI stage of 1 (indicating a less severe condition) might be more prone to experiencing side effects rather than benefits.
As expected, based on the second line chart in \autoref{fig: teaser}-B6, he found that as the lower limit of the AKI stage increased, the hazard ratio slightly decreased.
However, he said, \textit{``Although including patients with an AKI stage of 1 increases the hazard ratio, it remains at an acceptable level. Considering the obvious increase in the number of patients and the overall lower kidney risk in this group, it seems reasonable to set the lower limit of the AKI stage as 1.''} 
He applied the same approach to another criterion and concluded that the study should also include patients with a higher BMI. 

\textbf{The final decision.} 
\Eb discovered that the clinical trial would include over 5,000 patients after the exploration. Compared to the initial 1,000 patients, this represented a 5-fold increase in the number of patients. 
In addition, \Eb found that the other outcome metrics remained acceptable. 
\sidecomment{R2C8}
\revise{There was a slight increase in the hazard ratio from the initial value of 0.59 to 0.72. Although the two values are distant from 1, they are not excessively low, indicating a similar level and suggesting that the drug is effective in the selected population.}
% \xingbo{Notably, the upper bound of the 95\% confidence interval for the hazard ratio was 1.02, suggesting the result was not statistically significant.
% However, \Eb reasoned that this marginal inclusion of 1 in the confidence interval might not reflect the actual experimental outcome and did not warrant rejecting the criterion candidate.}
Furthermore, he found that the kidney risk ratio remained below 1, and there was a notable decrease in the liver risk ratio compared to the initial settings.  
Overall, \Eb expressed excitement about these insights, as they indicate a substantial number of patients without a significant increase in risk.

\section{Expert Interview}\label{expert_interview}
To further evaluate the effectiveness of our system, we conducted one-on-one interviews via Zoom with five clinicians (\Pa-\Pe) who had not participated in the design process of \textit{TrialCompass} and had never used our system.
\sidecomment{R1C2, R1C4}
\revise{These participants were recommended by experts from our earlier formative study, based on our inclusion criteria—namely, having at least three years of experience in clinical trial design and having completed at least one full trial cycle. To ensure objectivity, we did not allow the experts to contact the participants directly; instead, we reached out via email or phone. Additionally, participants came from different hospitals, enhancing the independence and generalizability of their feedback.
Among them, three were male and two were female clinicians, with an average age of 41 (ranging from 26 to 55). 
Their average experience in clinical trials was 9.4 years (ranging from 3 to 21 years).
All of them specialized in kidney-related fields, as we intended to use the sepsis-associated acute kidney injury (introduced in Case II in \autoref{case_two}), a common and severe illness, as their exploration scenario.}
As eligibility criteria, outcomes, and detailed characteristics represent fundamental components of eligibility criteria designs, our interview specifically investigated how the Criterion-Outcome Exploration View and Detailed Characteristic Exploration View can facilitate the exploration and decision-making processes.
We began the session by providing a 10-minute introduction to the research background. 
Following that, we demonstrated the system functionality and usage through Case I for 20 minutes. Next, the experts could familiarize themselves with the system for 15 minutes. Then, the experts utilized our system to adjust the eligibility criteria in Case II for 30 minutes. 
During the process, we asked the experts to think aloud, allowing us to record their audio during the exploration process and facilitate our subsequent analysis. 
% Additionally, our system automatically recorded the widget usage frequency for further analysis. 
Afterward, we conducted a 12-minute interview to gather the experts' feedback on using the system. Lastly, we invited the experts to complete a questionnaire to rate our system's usability, which took approximately three minutes.

\textbf{System Workflow.} We summarized how our two approaches (\ie, the knowledge-driven approach and the outcome-driven approach) facilitate their decision-making process, respectively.

$\diamond$ \textbf{\textit{The knowledge-driven approach.}} 
All the experts have emphasized a key advantage. 
The knowledge-driven approach allows them to obtain a precise understanding of the outcomes related to eligibility criteria that they are familiar with at first.
\Pe expressed, \textit{``Clinicians might have a general sense of eligibility criteria, but it may not be specific enough. Therefore, we desire to know its exact outcomes through the knowledge-driven approach, thus guiding further exploration.''}
Additionally, several experts also mentioned that they could avoid examining too many options in the Outcome View.
\Pc stated that he could significantly reduce the number of options in the Outcome View once he determined several criteria through his knowledge upfront.
Lastly, \Pb emphasized the coherence provided by the knowledge-driven approach. By adjusting the criteria with his expertise, he expressed greater confidence in maintaining consistency throughout the optimization process.

$\diamond$ \textbf{\textit{The outcome-driven approach.}} 
The most prominent advantage of adopting the outcome-driven approach is quickly exploring and determining multiple criteria.
Additionally, clinicians can fill their knowledge gaps regarding complex combinations among the eligibility criteria.
Furthermore, the outcome-driven approach can help uncover more optimal candidates. Initially, \Pa established two types of eligibility criteria: one for what he perceived as the worst-performing group and the other for the best-performing group. 
\Pa observed a hazard ratio of 0.79 with a sample size of 3541 in the first group, while the second group had a smaller sample size of 512 and a lower hazard ratio of 0.59. This discrepancy led him to consider the possibility of better candidates lying between these two extremes. However, managing multiple criteria proved challenging, even for experienced clinicians. Consequently, he adopted the outcome-driven approach to explore additional points along the spectrum.
Finally, we were surprised to discover that \Pd started the exploration process through the outcome-driven approach. She mentioned that she hoped to prevent her knowledge bias. \textit{``Clinicians might possess similar knowledge bases. If I aim to discover more promising designs for eligibility criteria, starting the exploration through the outcome-driven approach could be effective.''}

\textbf{Visual Designs and Interactions.}
All the experts praised the clear and user-friendly visual designs and interactions of our system. First, they highlighted the smooth combination of the three sub-views in the Criterion-outcome Exploration View, which allowed for easy criterion setting, candidate outcome examination, and exploration history tracking. 
\Pa found the interaction between the Criterion View and Outcome View to be highly suited to his needs. He expressed, \textit{``This exploration approach can also be extended to various stratified research in the medical field.''}
\Pa and \Pe both expressed high appreciation for the ability to create stages, as it allowed them to effectively organize and recall different explorations. 
We evaluated their perception of the system using the NASA Task Load Index \cite{Sandra2006Nasa}, a 7-point scale. 
First, the overall system design is not complex, as indicated by the average scores for mental demand (3.4), physical demand (2.4), effort (2.8), and frustration (1.8). 
However, the temporal demand score is 4.6.
This could be attributed to the iterative nature of the exploration process, where experts engage in repeated iterations.
Furthermore, the average performance score is 1.7 (with scores closer to 1 indicating a higher level of perceived performance perfection), indicating that the experts are highly confident of the final result obtained from our system.

\textbf{Suggestions.}
Experts offered several suggestions for further enhancements. First, they recommended highlighting regions in the Outcome View to make it easier to identify interesting candidates. Second, \Pe suggested providing greater flexibility in organizing exploration, such as enabling a hierarchical tree format. Lastly, several experts proposed automatically recommending potentially important criteria to ensure they are not overlooked due to knowledge limitations.
\section{Discussion}
\textbf{Design Implications.}
We have identified two important aspects during the system design process.
\sidecomment{R1C6}
\revise{First, our findings highlight the importance of supporting clinicians in systematically tracking their iterative exploration process. In various clinical scenarios, the decision-making process is always non-linear and exploratory, often requiring backtracking, hypothesis refinement, and contextual sense-making~\cite{Thierry2014Diagnostic, gladstone2012comparative}. While prior work in visual analytics~\cite{xu2020survey} has emphasized the value of provenance tracking and cognitive support, our work reinforces these findings. We observed that allowing clinicians to define their own reasoning stages and annotate their thought process—through note-taking and snapshotting—helps externalize reasoning, making it easier to revisit and replicate successful strategies while avoiding redundant exploration.} 
% Our findings show that it is crucial to help experts systematically track their iterative exploration process.
% In our scenario, taking notes for each exploration stage serves as an effective approach, allowing experts to document and replicate successful exploration strategies in the subsequent analysis. 
\revise{Another key implication is the need to support the adaptive integration of clinical expertise within exploratory decision workflows. We observed that clinicians often shift between knowledge-driven and outcome-driven strategies, depending on their evolving goals and domain understanding. This echoes findings from VBridge~\cite{Cheng2022VBridge}, where clinicians alternated between forward and backward analyses when collaborating with AI. Our results suggest that such hybrid reasoning patterns may also apply in broader, non-AI contexts. Therefore, future systems should therefore accommodate these fluid transitions and support flexible, mixed-strategy reasoning.}

\textbf{Generalizability.}
In this work, we focus on the design of eligibility criteria in clinical trials, a crucial step before participant enrollment. 
However, conducting a clinical trial involves multiple steps \cite{friedman2015fundamentals}.
For instance, clinicians are tasked with patient screening during participant enrollment to confirm that participants adhere to the trial's criteria. 
Furthermore, after enrollment, they need to conduct experimental procedures and monitor patient conditions. 
Finally, they will assess adverse events and interpret the final results.
Our system can also be utilized in several steps. 
For example, the Criterion-outcome Exploration View can help clinicians organize and analyze the final treatment effectiveness of different subgroups.
Beyond clinical trials, our system can also inspire other decision-making scenarios.
For example, our system integrates outcome metrics with temporal detailed characteristics.
This can be used for long-term investment allocation. 
Currently, various models have been developed to predict how much profit and risk investment will bring \cite{Gu2021Research}. They allow investors to simulate various investment allocation strategies and assess the potential outcomes. 
Investors also require detailed information (\eg, market context or current investment trend) with these outcome metrics during their decision-making process.
Additionally, the combination of the knowledge-driven and outcome-driven approaches can allow them to better leverage their prior domain knowledge as well as data-driven techniques.

\textbf{Scalability.}
First, scalability issues can arise in the Outcome View. 
When the number of criteria or possible adjustments for each criterion increases, the number of criterion candidates will grow rapidly.
\sidecomment{R1C3}
\revise{However, since experts' initial exploration focuses on understanding the overall distribution in the scatter plot, even with a high number of criterion candidates, they can explore based on distribution without needing to examine individual candidates. Thus, the candidate count does not hinder their exploration.
Additionally, our system allows clinicians to take an iterative approach. They can start by avoiding overly granular adjustments for each eligibility criterion. After gaining insight into the appropriate adjustment range, they can then fine-tune the criteria more precisely.
In the future, we plan to introduce a sampling approach. When data volumes are large, uniform sampling initially will not compromise users' judgment of the overall distribution. As users refine their selections, we can gradually reintroduce previously hidden candidates. This method can ensure smooth exploration while maintaining the system's rendering capabilities.}
Scalability issues may also arise in the Exploration View with numerous exploration records, making it difficult for experts to grasp the overall context. Currently, we allow clinicians to specify important stages. In the future, we could add features like hierarchical organization of exploration history.

\textbf{Limitations.}
Firstly, our system incorporated five outcome metrics based on our literature survey and expert interviews. 
\revise{While we expanded the metrics compared to previous tools, some suggested metrics were not included due to the dataset limitation. 
Incorporating more relevant metrics in the system can be a future work.
For example, detailed heart-related risk metrics could provide important insights into cardiovascular complications and help better assess patient safety during trials.}
\sidecomment{R3C3}
Secondly, our system is currently limited to examining kidney and liver details from EHR data. Clinicians may need to explore more detailed information and define a broader spectrum of risk events, which presents an opportunity for future enhancements.

\section{Conclusion}
In this work, we proposed \systemname, a visual analytics system to assist clinicians in designing eligibility criteria for clinical trials. We developed a novel workflow that enables exploration of eligibility criteria through both knowledge-driven and outcome-driven approaches. Additionally, we integrated a history-tracking feature to support clinicians in their iterative design process. Using the MIMIC IV dataset, we conducted expert interviews and case studies, uncovering new insights for eligibility criteria in two major diseases. 
Finally, we highlighted several research opportunities that arise from applying visualization techniques to enhance clinical trial workflows.

% \end{spacing}

% \input{section/10_supplementary.tex}
\acknowledgments{
	We sincerely thank all our collaborators and reviewers. We also appreciate experts who participated in the development and validation of our system. Lastly, we express our gratitude to all the partners who have helped us with our project.
}
% \balance
\bibliographystyle{abbrv-doi-hyperref}
\bibliography{ref}

@article{liu2021evaluating,
  title={{Evaluating Eligibility Criteria of Oncology Trials using Real-world Data and AI}},
  author={Liu, R. and Rizzo, S. and Whipple, S. et al.},
  journal={Nature},
  volume={592},
  number={7855},
  pages={629--633},
  year={2021},
  doi={10.1038/s41586-021-03430-5}
}

@article{fehrenbacher2009randomized,
  title={{Randomized Clinical Trial Eligibility Rates for Chemotherapy (CT) and Antiangiogenic Therapy (AAT) in a Population-based Cohort of Newly Diagnosed Non-small Cell Lung Cancer (NSCLC) Patients}},
  author={Fehrenbacher, L and Ackerson, L and Somkin, C},
  journal={Journal of Clinical Oncology},
  volume={27},
  number={15\_suppl},
  pages={6538--6538},
  year={2009},
  publisher={American Society of Clinical Oncology},
  doi = {10.1200/jco.2009.27.15_suppl.6538}
}

@article{huang2018clinical,
  title={{Clinical Trials Recruitment Planning: A Proposed Framework From the Clinical Trials Transformation Initiative}},
  author={Huang, Grant D and Bull, Jonca and McKee, Kelly Johnston and Mahon, Elizabeth and Harper, Beth and Roberts, Jamie N and CTTI Recruitment Project Team and others},
  journal={Contemporary Clinical Trials},
  volume={66},
  pages={74--79},
  year={2018},
  publisher={Elsevier},
  doi = {10.1016/j.cct.2018.01.003}
}

@article{kim2021towards,
  title={{Towards Clinical Data-driven Eligibility Criteria Optimization for Interventional COVID-19 Clinical Trials}},
  author={Kim, JH and Ta, CN and Liu, C and Sung, C and Butler, AM and Stewart, LA and Ena, L and Rogers, JR and Lee, J and Ostropolets, A and Ryan, PB and Liu, H and Lee, SM and Elkind, MSV and Weng, C},
  journal={Journal of the American Medical Informatics Association},
  volume={28},
  number={1},
  pages={14--22},
  year={2021},
  doi={10.1093/jamia/ocaa276}
}

@ARTICLE{zhao2017skylens,
  author={Zhao, Xun and Wu, Yanhong and Cui, Weiwei and Du, Xinnan and Chen, Yuan and Wang, Yong and Lee, Dik Lun and Qu, Huamin},
  journal={IEEE Transactions on Visualization and Computer Graphics}, 
  title={{SkyLens: Visual Analysis of Skyline on Multi-Dimensional Data}}, 
  year={2018},
  volume={24},
  number={1},
  pages={246--255},
  doi={10.1109/TVCG.2017.2744738}}

@article{kwee2023target,
    author = {Kwee, Sandi A. and Wong, Linda L. and Ludema, Christina and Deng, Chris K. and Taira, Deborah and Seto, Todd and Landsittel, Douglas},
    title = {{Target Trial Emulation: A Design Tool for Cancer Clinical Trials}},
    journal = {JCO Clinical Cancer Informatics},
    volume = {},
    number = {7},
    pages = {e2200140},
    year = {2023},
    doi = {10.1200/CCI.22.00140},
}

@inproceedings{weng2018homefinder,
    author = {Weng, Di and Zhu, Heming and Bao, Jie and Zheng, Yu and Wu, Yingcai},
    title = {{HomeFinder Revisited: Finding Ideal Homes with Reachability-Centric Multi-Criteria Decision Making}},
    year = {2018},
    isbn = {9781450356206},
    doi = {10.1145/3173574.3173821},
    booktitle = {Proceedings of the CHI Conference on Human Factors in Computing Systems},
    numpages = {12},
}

@ARTICLE{liu2017smartadp,
  author={Liu, Dongyu and Weng, Di and Li, Yuhong and Bao, Jie and Zheng, Yu and Qu, Huamin and Wu, Yingcai},
  journal={IEEE Transactions on Visualization and Computer Graphics}, 
  title={{SmartAdP: Visual Analytics of Large-scale Taxi Trajectories for Selecting Billboard Locations}}, 
  year={2017},
  volume={23},
  number={1},
  pages={1--10},
  doi={10.1109/TVCG.2016.2598432}
}

@article{chen2024fslens,
  author={Chen, Longfei and Wang, He and Ouyang, Yang and Zhou, Yang and Wang, Naiyu and Li, Quan},
  journal={IEEE Transactions on Visualization and Computer Graphics}, 
  title={{FSLens: A Visual Analytics Approach to Evaluating and Optimizing the Spatial Layout of Fire Stations}}, 
  year={2024},
  volume={30},
  pages={847--857},
  doi={10.1109/TVCG.2023.3327077}}

@article{Aggarwal2019StudyDP,
  title={{Study Designs: Part 4 – Interventional Studies}},
  author={Rakesh Aggarwal and Priya Ranganathan},
  year={2019},
  volume={10},
  number={3},
  pages={137--139},
  doi = {10.4103/picr.PICR_91_19},
  url={https://api.semanticscholar.org/CorpusID:196458381},
  journal={Perspectives in Clinical Research},
}

@article{ferreira2017types,  
  author={Ferreira, Juliana Carvalho and Patino, Cecilia Maria}, 
  title={{Types of Outcomes in Clinical Research}},  
  volume={43},  
  ISSN={1806--3713},  
  DOI={10.1590/S1806-37562017000000021},  
  number={1},  
  journal={Jornal Brasileiro de Pneumologia},  
  publisher={Sociedade Brasileira de Pneumologia e Tisiologia},  
  year={2017},  
  pages={5} 
}

@ARTICLE{Kwon2021DPVis,
  author={Kwon, Bum Chul and Anand, Vibha and Severson, Kristen A. and Ghosh, Soumya and Sun, Zhaonan and Frohnert, Brigitte I. and Lundgren, Markus and Ng, Kenney},
  journal={IEEE Transactions on Visualization and Computer Graphics}, 
  title={{DPVis: Visual Analytics With Hidden Markov Models for Disease Progression Pathways}}, 
  year={2021},
  volume={27},
  number={9},
  pages={3685--3700},
  doi={10.1109/TVCG.2020.2985689}}

@article{Wang2022ThreadStates,
  title={{ThreadStates: State-based Visual Analysis of Disease Progression}},
  author={Wang, Q and Mazor, T and Harbig, TA and Cerami, E and Gehlenborg, N},
  journal={IEEE Transactions on Visualization and Computer Graphics},
  volume={28},
  number={1},
  pages={238--247},
  year={2022},
  publisher={IEEE},
  doi={10.1109/TVCG.2021.3114840},
}

@article{Sultanum2023ChartWalk,
  title={{ChartWalk: Navigating Large Collections of Text Notes in Electronic Health Records for Clinical Chart Review}},
  author={Sultanum, N and Naeem, F and Brudno, M and Chevalier, F},
  journal={IEEE Transactions on Visualization and Computer Graphics},
  volume={29},
  number={1},
  pages={1244--1254},
  year={2023},
  publisher={IEEE},
  doi={10.1109/TVCG.2022.3209444},
}

@ARTICLE {Linhares2023ClinicalPath,
    author = {C. G. Linhares and D. M. Lima and J. R. Ponciano and M. M. Olivatto and M. A. Gutierrez and J. Poco and C. Traina and A. M. Traina},
    journal = {IEEE Transactions on Visualization and Computer Graphics},
    title = {{ClinicalPath: A Visualization Tool to Improve the Evaluation of Electronic Health Records in Clinical Decision-Making}},
    year = {2023},
    volume = {29},
    number = {10},
    pages = {4031--4046},
    doi = {10.1109/TVCG.2022.3175626},
    publisher = {IEEE Computer Society},
}

@article{Cheng2022VBridge,
  title={{VBridge: Connecting the Dots Between Features and Data to Explain Healthcare Models}},
  author={Cheng, F and Liu, D and Du, F and Lin, Y and Zytek, A and Li, H and Qu, H and Veeramachaneni, K},
  journal={IEEE Transactions on Visualization and Computer Graphics},
  volume={28},
  number={1},
  pages={378--388},
  year={2022},
  publisher={IEEE},
  doi={10.1109/TVCG.2021.3114836}
}

@article{Desai2020recruitment,
    Title = {{Recruitment and Retention of Participants in Clinical Studies: Critical Issues and Challenges}},
    Author = {Desai, Mira},
    DOI = {10.4103/picr.picr_6_20},
    Number = {2},
    Volume = {11},
    Year = {2020},
    Journal = {Perspectives in Clinical Research},
    ISSN = {2229--3485},
    Pages = {51--53},
}

@article {Royston2013Restricted,
    Title = {{Restricted Mean Survival Time: An Alternative to the Hazard Ratio for the Design and Analysis of Randomized Trials with a Time-to-event Outcome}},
    Author = {Royston, Patrick and Parmar, Mahesh K B},
    DOI = {10.1186/1471-2288-13-152},
    Volume = {13},
    Year = {2013},
    Journal = {BMC Medical Research Methodology},
    Pages = {152},
}

@article{Kelsey2022Inclusion,
    title = {{Inclusion and Diversity in Clinical Trials: Actionable Steps to Drive Lasting Change}},
    journal = {Contemporary Clinical Trials},
    volume = {116},
    pages = {106740},
    year = {2022},
    doi = {https://doi.org/10.1016/j.cct.2022.106740},
    author = {Michelle D. Kelsey and Bray Patrick-Lake and Raolat Abdulai and Uli C. Broedl and Adam Brown and Elizabeth Cohn and Lesley H. Curtis and Chris Komelasky and Michael Mbagwu and George A. Mensah and Robert J. Mentz and Amesika Nyaku and Stephanie O. Omokaro and Judy Sewards and Kendal Whitlock and Xinzhi Zhang and Gerald S. Bloomfield},
}

@ARTICLE{Jiang2024HealthPrism,
  author={Jiang, Zhihan and Chen, Handi and Zhou, Rui and Deng, Jing and Zhang, Xinchen and Zhao, Running and Xie, Cong and Wang, Yifang and Ngai, Edith C.H.},
  journal={IEEE Transactions on Visualization and Computer Graphics}, 
  title={{HealthPrism: A Visual Analytics System for Exploring Children's Physical and Mental Health Profiles with Multimodal Data}}, 
  year={2024},
  volume={30},
  number={1},
  pages={1205-1215},
  doi={10.1109/TVCG.2023.3326943}
}

@inproceedings{li2023trialview,
	Title = {TrialView: An AI-powered Visual Analytics System for Temporal Event Data in Clinical Trials},
	Author = {Li, Zuotian and Liu, Xiang and Cheng, Zelei and Chen, Yingjie and Tu, Wanzhu and Su, Jing},
	Year = {2024},
	booktitle = {Proceedings of the Hawaii International Conference on System Sciences},
	Pages = {1169—1178},
        doi={
https://doi.org/10.48550/arXiv.2310.04586}
}

@article{spotswood2004hazard,
    author = {Spotswood L. Spruance and Julia E. Reid and Michael Grace and Matthew Samore},
    title = {{Hazard Ratio in Clinical Trials}},
    journal = {Antimicrobial Agents and Chemotherapy},
    volume = {48},
    number = {8},
    pages = {2787--2792},
    year = {2004},
    doi = {10.1128/aac.48.8.2787-2792.2004},
}

@article{Pratt2000Evaluation,
    Title = {{Evaluation of Abnormal Liver-Enzyme Results in Asymptomatic Patients}},
    Author = {Pratt, DS and Kaplan, MM},
    DOI = {10.1056/nejm200004273421707},
    Number = {17},
    Volume = {342},
    Year = {2000},
    Journal = {The New England Journal of Medicine},
    Pages = {1266—1271},
}

@article {Bostom2002Predictive,
    Title = {{Predictive Performance of Renal Function Equations for Patients with Chronic Kidney Disease and Normal Serum Creatinine Levels}},
    Author = {Bostom, Andrew G and Kronenberg, Florian and Ritz, Eberhard},
    DOI = {10.1097/01.asn.0000022011.35035.f3},
    Volume = {13},
    Year = {2002},
    number = {8},
    Journal = {American Society of Nephrology},
    ISSN = {1046-6673},
    Pages = {2140--2144},
}

@article{Dziura2013Strategies,
    author={Dziura,James D. and Post,Lori A. and Zhao,Qing and Fu,Zhixuan and Peduzzi,Peter},
    year={2013},
    title={{Strategies for Dealing with Missing Data in Clinical Trials: From Design to Analysis}},
    journal={Yale Journal of Biology and Medicine},
    volume={86},
    number={3},
    pages={343-358},
    doi={10.4103/cmi.cmi_8_24}
}

@book{blass2015basic,
  title={{Basic Principles of Drug Discovery and Development}},
  author={Blass, Benjamin E},
  year={2015},
  publisher={Elsevier},
  doi={https://doi.org/10.1016/C2017-0-02030-X}
}

@article{rosenbaum1983propensity,
    author = {Rosenbaum, Paul R. and Rubin, Donald B.},
    title = {{The Central Role of the Propensity Score in Observational Studies for Causal Effects}},
    journal = {Biometrika},
    volume = {70},
    number = {1},
    pages = {41-55},
    year = {1983},
    doi = {10.1093/biomet/70.1.41},
}

@article{Austin2011Optimal,
    Title = {{Optimal Caliper Widths for Propensity-score Matching When Estimating Differences in Means and Differences in Proportions in Observational Studies}},
    Author = {Austin, Peter C},
    Year = {2011},
    DOI = {10.1002/pst.433},
    Volume = {10},
    number = {2},
    Journal = {Pharmaceutical Statistics},
    ISSN = {1539-1604},
    Pages = {150--161},
}

@article{Zhang2019IDMVis,
    author = {Zhang, Yixuan and Chanana, Kartik and Dunne, Cody},
    title = {{IDMVis: Temporal Event Sequence Visualization for Type 1 Diabetes Treatment Decision Support}},
    year = {2019},
    publisher = {IEEE Educational Activities Department},
    volume = {25},
    number = {1},
    issn = {1077-2626},
    doi = {10.1109/TVCG.2018.2865076},
    journal = {IEEE Transactions on Visualization and Computer Graphics},
    pages = {512--522},
}

@ARTICLE{Kwon2019RetainVis,
  author={Kwon, Bum Chul and Choi, Min-Je and Kim, Joanne Taery and Choi, Edward and Kim, Young Bin and Kwon, Soonwook and Sun, Jimeng and Choo, Jaegul},
  journal={IEEE Transactions on Visualization and Computer Graphics}, 
  title={{RetainVis: Visual Analytics with Interpretable and Interactive Recurrent Neural Networks on Electronic Medical Records}}, 
  year={2019},
  volume={25},
  number={1},
  pages={299--309},
  doi={10.1109/TVCG.2018.2865027}}

@ARTICLE{Weng2019SRVis,
  author={Weng, Di and Chen, Ran and Deng, Zikun and Wu, Feiran and Chen, Jingmin and Wu, Yingcai},
  journal={IEEE Transactions on Visualization and Computer Graphics}, 
  title={{SRVis: Towards Better Spatial Integration in Ranking Visualization}}, 
  year={2019},
  volume={25},
  number={1},
  pages={459--469},
  doi={10.1109/TVCG.2018.2865126}}

@inproceedings{Carenini2004ValueCharts,
    author = {Carenini, Giuseppe and Loyd, John},
    title = {{ValueCharts: Analyzing Linear Models Expressing Preferences and Evaluations}},
    year = {2004},
    isbn = {1581138679},
    doi = {10.1145/989863.989885},
    booktitle = {Proceedings of the Working Conference on Advanced Visual Interfaces},
    pages = {150--157},
}

@ARTICLE{Wall2018Podium,
  author={Wall, Emily and Das, Subhajit and Chawla, Ravish and Kalidindi, Bharath and Brown, Eli T. and Endert, Alex},
  journal={IEEE Transactions on Visualization and Computer Graphics}, 
  title={{Podium: Ranking Data Using Mixed-Initiative Visual Analytics}}, 
  year={2018},
  volume={24},
  number={1},
  pages={288--297},
  doi={10.1109/TVCG.2017.2745078}}

@ARTICLE{Gratzl2013LineUp,
  author={Gratzl, Samuel and Lex, Alexander and Gehlenborg, Nils and Pfister, Hanspeter and Streit, Marc},
  journal={IEEE Transactions on Visualization and Computer Graphics}, 
  title={{LineUp: Visual Analysis of Multi-Attribute Rankings}}, 
  year={2013},
  volume={19},
  number={12},
  pages={2277-2286},
  doi={10.1109/TVCG.2013.173}}

@ARTICLE{Pajer2017WeightLifter,
  author={Pajer, Stephan and Streit, Marc and Torsney-Weir, Thomas and Spechtenhauser, Florian and Möller, Torsten and Piringer, Harald},
  journal={IEEE Transactions on Visualization and Computer Graphics}, 
  title={{WeightLifter: Visual Weight Space Exploration for Multi-Criteria Decision Making}}, 
  year={2017},
  volume={23},
  number={1},
  pages={611--620},
  doi={10.1109/TVCG.2016.2598589}}

@article{Claessens2013Are,
  AUTHOR = {Claessens, Yann-Erick and Aegerter, Philippe and Boubaker, Hamdi and Guidet, Bertrand and Cariou, Alain},
  JOURNAL = {{Critical Care}},
  PUBLISHER = {{BioMed Central}},
  VOLUME = {17},
  NUMBER = {3},
  ARTICLENO = {R89},
  NUMPAGES = {9},
  YEAR = {2013},
  DOI = {10.1186/cc12734}
}

@book{friedman2015fundamentals,
  title={Fundamentals of Clinical Trials},
  author={Friedman, Lawrence M and Furberg, Curt D and DeMets, David L and Reboussin, David M and Granger, Christopher B},
  year={2015},
  publisher={Springer},
  doi={https://doi.org/10.5213/inj.2013.17.2.96}
}

@INPROCEEDINGS{Debek2017timeline,
  author={Dabek, Filip and Jimenez, Elizabeth and Caban, Jesus J.},
  booktitle={2017 IEEE Workshop on Visual Analytics in Healthcare}, 
  title={{A Timeline-based Framework for Aggregating and Summarizing Electronic Health Records}}, 
  year={2017},
  volume={},
  number={},
  pages={55--61},
  doi={10.1109/VAHC.2017.8387501}}

@inproceedings{Faiola2011Advancing,
    author = {Faiola, Anthony and Newlon, Chris},
    title = {{Advancing Critical Care in the ICU: A Human-centered Biomedical Data Visualization Systems}},
    year = {2011},
    isbn = {9783642217159},
    booktitle = {Proceedings of the International Conference on Ergonomics and Health Aspects of Work with Computers},
    pages = {119--128},
    doi = {10.1007/978-3-642-21716-6_13}
}

@article{Hirsch2014HARVEST,
    author = {Hirsch, Jamie S and Tanenbaum, Jessica S and Lipsky Gorman, Sharon and Liu, Connie and Schmitz, Eric and Hashorva, Dritan and Ervits, Artem and Vawdrey, David and Sturm, Marc and Elhadad, Noémie},
    title = {{HARVEST: A longitudinal Patient Record Summarizer}},
    journal = {Journal of the American Medical Informatics Association},
    volume = {22},
    number = {2},
    pages = {263--274},
    year = {2014},
    doi = {10.1136/amiajnl-2014-002945},
}

@article{Sandra2006Nasa,
    author = {Sandra G. Hart},
    title ={{Nasa-Task Load Index (NASA-TLX); 20 Years Later}},
    journal = {Human Factors and Ergonomics Society Annual Meeting},
    volume = {50},
    number = {9},
    pages = {904--908},
    year = {2006},
    doi = {10.1177/154193120605000909},
}

@article{Opal2013Effect,
    author = {Opal, Steven M. and Laterre, Pierre-Francois and Francois, Bruno and LaRosa, Steven P. and Angus, Derek C. and Mira, Jean-Paul and Wittebole, Xavier and Dugernier, Thierry and Perrotin, Dominique and Tidswell, Mark and Jauregui, Luis and Krell, Kenneth and Pachl, Jan and Takahashi, Takeshi and Peckelsen, Claus and Cordasco, Edward and Chang, Chia-Sheng and Oeyen, Sandra and Aikawa, Naoki and Maruyama, Tatsuya and Schein, Roland and Kalil, Andre C. and Van Nuffelen, Marc and Lynn, Melvyn and Rossignol, Daniel P. and Gogate, Jagadish and Roberts, Mary B. and Wheeler, Janice L. and Vincent, Jean-Louis and ACCESS Study Group, for the},
    title = {{Effect of Eritoran, an Antagonist of MD2-TLR4, on Mortality in Patients With Severe Sepsis: The ACCESS Randomized Trial}},
    journal = {Journal of the American Medical Informatics Association},
    volume = {309},
    number = {11},
    pages = {1154-1162},
    year = {2013},
    month = {03},
    doi = {10.1001/jama.2013.2194} 
}

@article{Berk2020Effect,
    author = {Berk, Michael and Woods, Robyn L. and Nelson, Mark R. and Shah, Raj C. and Reid, Christopher M. and Storey, Elsdon and Fitzgerald, Sharyn and Lockery, Jessica E. and Wolfe, Rory and Mohebbi, Mohammadreza and Dodd, Seetal and Murray, Anne M. and Stocks, Nigel and Fitzgerald, Paul B. and Mazza, Catherine and Agustini, Bruno and McNeil, John J.},
    title = {{Effect of Aspirin vs Placebo on the Prevention of Depression in Older People: A Randomized Clinical Trial}},
    journal = {JAMA Psychiatry},
    volume = {77},
    number = {10},
    pages = {1012-1020},
    year = {2020},
    doi = {10.1001/jamapsychiatry.2020.1214},
    url = {https://doi.org/10.1001/jamapsychiatry.2020.1214},
}

@article{Fang2023data,
    title = {{A Data-driven Approach to Optimizing Clinical Study Eligibility Criteria}},
    journal = {Journal of Biomedical Informatics},
    volume = {142},
    pages = {104375},
    year = {2023},
    doi = {https://doi.org/10.1016/j.jbi.2023.104375},
    author = {Yilu Fang and Hao Liu and Betina Idnay and Casey Ta and Karen Marder and Chunhua Weng}
}

@article{Gu2021Research,
    author = {Gu, Cong},
    title = {{Research on Prediction of Investment Fund’s Performance before and after Investment Based on Improved Neural Network Algorithm}},
    journal = {Wireless Communications and Mobile Computing},
    volume = {2021},
    number = {1},
    pages = {5519213},
    doi = {https://doi.org/10.1155/2021/5519213},
    year = {2021}
}

@article{Ettore2016Adverse,
    author = {Ettore Bartoli},
    title = {{Adverse Effects of Drugs on the Kidney}},
    journal = {European Journal of Internal Medicine},
    volume = {28},
    pages = {1-8},
    year = {2016},
    doi = {https://doi.org/10.1016/j.ejim.2015.12.001},
}

@article{Anna2016Adverse,
    author = {Anna Licata},
    title = {{Adverse Drug Reactions and Organ Damage: The Liver}},
    journal = {European Journal of Internal Medicine},
    volume = {28},
    pages = {9-16},
    year = {2016},
    doi = {https://doi.org/10.1016/j.ejim.2015.12.017}
}

@article{Wang2022EHRSTAR,
	title = {{EHR STAR: The State-Of-the-Art in Interactive EHR Visualization}},
	doi = {10.1111/cgf.14424},
	issue = {1},
	journal = {Computer Graphics Forum},
	pages = {69-105},
	url = {https://nottingham-repository.worktribe.com/output/6348971},
	volume = {41},
	year = {2022},
	author = {Wang, Qiru and Laramee, R. S.}
}

@article{johnson2023mimic,
  title={{MIMIC-IV, A Freely Accessible Electronic Health Record Dataset}},
  author={Johnson, Alistair EW and Bulgarelli, Lucas and Shen, Lu and Gayles, Alvin and Shammout, Ayad and Horng, Steven and Pollard, Tom J and Hao, Sicheng and Moody, Benjamin and Gow, Brian and others},
  journal={Scientific Data},
  volume={10},
  number={1},
  pages={1},
  year={2023},
  doi={https://doi.org/10.1038/s41597-022-01899-x}
}

@INPROCEEDINGS{Kuo2023Animal,
  author={Kuo, Yun-Hsin and Martínez-López, Beatriz and Ma, Kwan-Liu},
  booktitle={IEEE Pacific Visualization Symposium}, 
  title={{Investigating Animal Infectious Diseases with Visual Analytics}}, 
  year={2023},
  volume={},
  number={},
  pages={71-81},
  doi={10.1109/PacificVis56936.2023.00015}}

@article{Thierry2014Diagnostic,
title = {{How and When Do Expert Emergency Physicians Generate and Evaluate Diagnostic Hypotheses? A Qualitative Study Using Head-Mounted Video Cued-Recall Interviews}},
journal = {Annals of Emergency Medicine},
volume = {64},
number = {6},
pages = {575-585},
year = {2014},
doi = {https://doi.org/10.1016/j.annemergmed.2014.05.003},
author = {Thierry Pelaccia and Jacques Tardif and Emmanuel Triby and Christine Ammirati and Catherine Bertrand and Valérie Dory and Bernard Charlin}
}

@article{gladstone2012comparative,
  title={{Comparative Theories in Clinical Decision Making and Their Application to Practice: A Reflective Case Study}},
  author={Gladstone, Nicholas},
  journal={British Journal of Anaesthetic \& Recovery Nursing},
  volume={13},
  number={3-4},
  pages={65--71},
  year={2012},
  publisher={Cambridge University Press},
  doi={10.1017/S1742645612000435}
}

@inproceedings{xu2020survey,
  title={{Survey on the Analysis of User Interactions and Visualization Provenance}},
  author={Xu, Kai and Ottley, Alvitta and Walchshofer, Conny and Streit, Marc and Chang, Remco and Wenskovitch, John},
  booktitle={Computer Graphics Forum},
  volume={39},
  number={3},
  pages={757--783},
  year={2020},
  doi={https://doi.org/10.1111/cgf.14035}
}

@article{Braun2006Thematic,
    author={Braun, Virginia and Clarke, Victoria},
    year={2006},
    title={{Using Thematic Analysis in Psychology}},
    journal={Qualitative Research in Psychology},
    volume={3},
    number={2},
    pages={77-101},
    isbn={14780887},
    doi = {10.1191/1478088706qp063oa},
}

@article{sheng2025design,
  title={Design Patterns of Human-AI Interfaces in Healthcare},
  author={Sheng, Rui and Shi, Chuhan and Lotfi, Sobhan and Liu, Shiyi and Perer, Adam and Qu, Huamin and Cheng, Furui},
  journal={arXiv preprint arXiv:2507.12721},
  year={2025},
  doi={10.48550/arXiv.2507.12721}
}

@ARTICLE{Wang2025SynthLens,
  author={Wang, Qipeng and Sheng, Rui and Ruan, Shaolun and Jin, Xiaofu and Shi, Chuhan and Zhu, Min},
  journal={IEEE Transactions on Visualization and Computer Graphics}, 
  title={SynthLens: Visual Analytics for Facilitating Multi-step Synthetic Route Design}, 
  year={2025},
  volume={},
  number={},
  pages={1-14},
  doi={10.1109/TVCG.2025.3552134}}
\clearpage
\appendix
\renewcommand{\thesection}{\Alph{section}}
\end{document}